\documentclass[11pt]{article}
\pdfoutput=1

\usepackage[margin=0.75in]{geometry}

\usepackage[ascii]{inputenc}
\usepackage[bookmarksnumbered,hypertexnames=false,colorlinks=true,linkcolor=blue,urlcolor=blue,citecolor=blue,anchorcolor=green]{hyperref}
\usepackage{bbm,braket,microtype,mathrsfs,amsmath,amssymb,color,amsthm,mathtools,fullpage,graphicx,enumitem,ae,aecompl,float,mathabx,caption,bm,mathdots}
\usepackage[capitalize]{cleveref}
\usepackage[numbers,sort&compress]{natbib}
\usepackage[T1]{fontenc}
\usepackage[english]{babel}
\usepackage{newpxtext}
\usepackage{thmtools,thm-restate}
\usepackage{xpatch}
\DeclareMathOperator{\tr}{tr}
\DeclareMathOperator{\diag}{diag}
\DeclareMathOperator{\SL}{SL}
\DeclareMathOperator{\T}{T}
\DeclareMathOperator{\ST}{ST}
\DeclareMathOperator{\GL}{GL}
\DeclareMathOperator{\Ten}{Ten}

\DeclareMathOperator{\Det}{Det}

\DeclareMathOperator{\poly}{poly}

\DeclareMathOperator{\srk}{slice-rank}

\DeclareMathOperator{\conv}{conv}

\newtheorem{thm}{Theorem}[section]\crefname{thm}{Theorem}{Theorems}
\newtheorem{lem}[thm]{Lemma}\crefname{lem}{Lemma}{Lemmas}
\newtheorem{prb}[thm]{Problem}\crefname{prb}{Problem}{Problems}
\newtheorem{rem}[thm]{Remark}\crefname{rem}{Remark}{Remarks}
\newtheorem{cor}[thm]{Corollary}\crefname{cor}{Corollary}{Corollaries}
\newtheorem{dfn}[thm]{Definition}\crefname{dfn}{Definition}{Definitions}
\newtheorem{prp}[thm]{Proposition}\crefname{prp}{Proposition}{Propositions}
\crefname{exa}{Example}{Examples}
\floatstyle{boxed}\newfloat{Algorithm}{ht}{alg}\crefname{Algorithm}{Algorithm}{Algorithms}

\floatstyle{boxed}\newfloat{ Algorithm}{ht}{algn}\crefname{algn}{Algorithm}{Algorithms}

\renewcommand{\vec}{\bm}
\newcommand{\CC}{\mathbb C}
\newcommand{\FF}{\mathbb F}
\newcommand{\RR}{\mathbb R}

\newcommand{\ZZ}{\mathbb Z}
\newcommand{\NN}{\mathbb N}
\newcommand{\QQ}{\mathbb Q}
\newcommand{\ot}{\otimes}

\newcommand{\eps}{\varepsilon}

\newcommand{\norm}[1]{\lVert#1\rVert}
\newcommand{\Norm}[1]{\left\lVert#1\right\rVert}

\newcommand{\cO}{\mathcal{O}}
\newcommand{\clO}{\overline{\cO}}
\newcommand{\cN}{\mathcal{N}}
\newcommand{\cM}{\mathcal{M}}

\newcommand{\perm}{\textnormal{perm}}
\newcommand{\ds}{\textnormal{ds}}
\newcommand{\mat}{\textnormal{Mat}}
\newcommand{\supp}{\textnormal{supp}}

\newcommand{\cP}{\mathsf{P}}

\newcommand{\csP}{\mathsf{\# P}}

\newcommand{\bB}{\mathbf{B}}
\newcommand{\bp}{\mathbf{p}}

\begin{document}

\title{Recent progress on scaling algorithms and applications}

\author{
Ankit Garg \\
\texttt{Microsoft Research India} \\
\texttt{\href{mailto:garga@microsoft.com}{garga@microsoft.com}}
\and
Rafael Oliveira \\
\texttt{University of Toronto} \\
\texttt{\href{mailto:rafael@cs.toronto.edu}{rafael@cs.toronto.edu}}
}
\pagenumbering{gobble}
\maketitle
\begin{abstract}

Scaling problems have a rich and diverse history, and thereby have found numerous applications in several fields 
of science and engineering. For instance, the matrix scaling problem has had applications ranging from 
theoretical computer science to telephone forecasting, economics, statistics, optimization, among many other fields. 
Recently, a generalization of matrix scaling known as operator scaling has found applications
in non-commutative algebra, invariant theory, combinatorics and algebraic complexity; and a further generalization 
(tensor scaling) has found more applications in quantum information theory, geometric complexity theory 
and invariant theory.


In this survey, we will describe in detail the scaling problems mentioned above, showing how alternating minimization
algorithms naturally arise in this setting, and we shall present a general framework to rigorously 
analyze such algorithms. These simple problems and algorithms are not just 
applicable to diverse mathematical and CS areas, but also serve to bring out deep connections between them. As this framework makes extensive use of concepts from invariant theory, we also 
provide a very gentle introduction to basic concepts of invariant theory and how they are used to analyze alternating minimization algorithms for the scaling problems.

This survey is intended for a general computer science audience, and the only
background required is basic knowledge of calculus and linear algebra, thereby making it accessible to 
graduate students and even to advanced undergraduates.

\end{abstract}

\newpage
\tableofcontents
\newpage
\pagenumbering{arabic}

\section{Introduction}\label{sec:intro}

Scaling problems have been in the background of many important developments 
in theoretical computer science, often times in an implicit or disguised way. For instance, Forster's lower bound
on the sign-rank of a matrix~\cite{F02}, Linial et al.'s deterministic approximation of the permanent~\cite{LSW} and
quantitative versions of the Sylvester-Gallai theorem~\cite{BDWY11} are results which can be seen
as particular instances or consequences of scaling problems

Outside of computer science, scaling algorithms have appeared (implicitly and explicitly) in many different
areas, such as economics~\cite{S62}, statistics~\cite{Sink}, optimization~\cite{ROTHBLUM1989737}, telephone
forecasting~\cite{K37}, non-commutative algebra~\cite{garg2016deterministic}, 
functional analysis~\cite{garg2017algorithmic}, quantum information theory~\cite{gurvits2004classical} 
and many others.

When trying to solve a scaling problem, a natural alternating minimization algorithm comes to mind, and as such
these algorithms have been proposed independently by many researchers. The analysis of such alternating minimization
algorithms, on the other hand, has been a difficult task, with many different approaches being proposed for 
each scaling problem, and before recent works, without a unified way of analyzing such scaling algorithms.
In this survey, we exposit a unified way of analyzing the natural alternating minimization algorithms, 
which is based on the series of works~\cite{LSW, gurvits2004classical, garg2016deterministic, 
burgisser2017alternating, burgisser2018efficient}.

This framework is a welcome addition to the rich and ever growing theory of optimization. The contributions to the theory of optimization are multifold. First is providing a general framework in which a general optimization heuristic, alternating minimization, converges in a polynomial number of iterations. Secondly, the underlying optimization problems are non-convex and yet they can be solved efficiently.\footnote{The underlying problems are geodesically convex i.e. convex in a different geometry. We will not discuss geodesic convexity in this survey but say that it is an active area of research and these scaling problems provide several interesting challenges and applications for this area.} So the framework provides a seemingly new tool looking for applications. Thirdly, these scaling problems give rise to a rich class of polytopes, called moment polytopes, which have exponentially many facets and yet there exist weak membership (conjecturally strong membership) oracles for them (see~\cite{garg2017algorithmic, burgisser2018efficient}). It remains to be seen if these moment polytopes can capture a large class of combinatorial polytopes (some of them they already can - \cite{garg2017algorithmic}).

This survey is divided as follows: in 
Section~\ref{sec:scaling_algs}, we formally describe the three
scaling problems that we study, together with the natural alternating minimization algorithms proposed for them. 
In Section~\ref{sec:inv-theory} we give an elementary introduction to invariant theory, with many examples, 
discussing how the scaling problems defined in Section~\ref{sec:scaling_algs} are particular instances 
of more general invariant theory questions. In Section~\ref{sec:analysis}, we provide a unified, 3-step analysis
of the alternating minimization algorithms proposed in Section~\ref{sec:scaling_algs}, showing how invariant 
theory is used in the analysis of such algorithms. In Section~\ref{sec:apps}, we give a detailed discussion
of some of the numerous applications of scaling algorithms, providing more references for the interested reader.
In Section~\ref{sec:conclusion} we conclude this survey, presenting further directions and future work to be done
in the area, and in Section~\ref{sec:related-works} we discuss related works (old and new) which we could not cover
in this survey due to space and time constraints, but otherwise would perfectly fit within the scope of the survey.

\section{Scaling: problems and algorithms}\label{sec:scaling_algs}
We first describe the various scaling problems and natural algorithms for them based on alternating minimization. \cref{subsec:matrix_scaling} studies matrix scaling, \cref{subsec:op_scaling} studies operator scaling and \cref{subsec:tensor_scaling} studies tensor scaling.

\subsection{Matrix scaling}\label{subsec:matrix_scaling}

The simplest scaling problem is matrix scaling, which dates back to Kruithoff~\cite{K37} in telephone forecasting 
and Sinkhorn~\cite{Sink} in statistics. There is a huge body of literature on this problem (see for 
instance~\cite{ROTHBLUM1989737, KALANTARI199687, LSW, GurYianilos, I16, AllenZhuLOW17, CohenMTV17} and references 
therein). In this subsection we will describe Sinkhorn's algorithm and the analysis of it done in~\cite{LSW}. 
We refer the reader to the last two references above for more sophisticated algorithms for matrix scaling.

We start with a few definitions, the first being the definition of a scaling of a matrix.

\begin{dfn}[Scaling of a matrix]\label{dfn:scaling_matrix} Suppose we are given an $n \times n$ non-negative 
(real) matrix $A$. We say that $A'$ is a {\em scaling} of $A$ if it can be obtained by multiplying the rows and 
columns of $A$ by positive scalars. In other words, $A'$ is a scaling of $A$ if there exist diagonal matrices $B,C$ 
(with positive entries) s.t. $A' = B A C$.
\end{dfn}

Next is the definition of a doubly stochastic matrix.

\begin{dfn}[Doubly stochastic]\label{dfn:doubly_stochastic}
An $n \times n$ non-negative matrix is said to be doubly stochastic if all of its row and column sums are equal to $1$.
\end{dfn}

The matrix scaling problem is simple to describe: given an $n \times n$ non-negative matrix $A$, find a scaling of $A$ which is doubly stochastic (if one exists). It turns out that an approximate version of the problem is more natural and has more structure, and as we will see in Section~\ref{sec:inv-theory}, this is not by accident. To define the approximate version, we will need another definition which is a quantification of how close a matrix is to being doubly stochastic.

\begin{dfn}\label{dfn:ds_matrix}
Given an $n \times n$ non-negative matrix $A$, define its {\em distance to doubly stochastic} to be  
$$
\ds(A) = \sum_{i=1}^n (r_i - 1)^2 + \sum_{j=1}^n (c_j - 1)^2
$$
Where $r_i, c_j$ denote the $i^{\text{th}}$ row and $j^{\text{th}}$ column sums of $A$, respectively.
\end{dfn}

With this notion of distance we can define the $\eps$-scaling problem, whose goal is to find a scaling $A'$ of 
$A$ s.t. $\ds(A') \le \eps$ (if one exists). 

\begin{dfn}[Scalability]\label{dfn:scalability_matrix} We will say that a non-negative matrix $A$ is scalable if 
for all $\eps > 0$, there exists a scaling $A'_{\eps}$ of $A$ s.t. $\ds(A'_{\eps}) \le \eps$.
\end{dfn}

Given this definition, several natural questions arise. When is a matrix scalable? If a matrix is scalable, can 
one efficiently find an $\eps$-scaling? It turns out that answers to both questions are extremely pleasing! 
The answer to the first question is given by the following theorem (e.g. see \cite{ROTHBLUM1989737}).

\begin{thm}\label{thm:scalability_matrix} An $n \times n$ non-negative matrix $A$ is scalable iff $\perm(A) > 0$.\footnote{Here $\perm(A) = \sum_{\sigma \in S_n} \prod_{i=1}^n A_{i, \sigma}$ is the permanent of the matrix $A$.} In other words, $A$ is scalable iff the bipartite graph defined by the support of $A$ has a perfect matching.
\end{thm}

Learning this nice structural result, one is naturally lead to the second (algorithmic) question: If $A$ is 
scalable, how to efficiently find and $\eps$-scaling?\footnote{Historically, the quest for an algorithmic solution
to this problem preceded the structural results.} Towards this Sinkhorn \cite{Sink} suggested an 
extremely natural algorithm which was analyzed in \cite{LSW}.

\begin{Algorithm}[th!]
\textbf{Input}:
$n \times n$ non-negative matrix $A$ with rational entries of bit complexity at most $b$, and a distance
parameter $\eps > 0$.
\\[.1ex]

\textbf{Output:}
Either the algorithm correctly identifies that $A$ is not scalable, or it outputs non-negative diagonal matrices $B, C$ s.t. $\ds(B A C) \le \eps$.
\\[.1ex]

\textbf{Algorithm:}\vspace{-.2cm}

\begin{enumerate}
\item Initialize $B = C = I$.
\item If $A$ has an all-zero row or column, output not scalable and halt.
\item For $T$ iterations, apply the following procedure:
\begin{itemize}
\item If $\ds(B A C) \le \eps$, output $B, C$ and halt. 
\subitem Else, if $\sum_{i=1}^n (r_i - 1)^2 > \eps/2$,\footnotemark \ normalize the rows, that is, set  
$$B \leftarrow \diag(r_1^{-1},\ldots, r_n^{-1}) B$$ 
\subitem Otherwise normalize the columns.
\end{itemize}
\item Output that $A$ is not scalable.
\end{enumerate}

\caption{Sinkhorn's algorithm}\label{alg:matrix-scaling}
\end{Algorithm}
\footnotetext{$r_1,\ldots, r_n$ denote the row sums of the current matrix i.e. $BAC$.}


\begin{thm}[\cite{LSW}]\label{thm:matrix-scaling} \cref{alg:matrix-scaling} with $T = O(n(b+\log(n))/\eps)$ iterations works correctly. That is, if the algorithm outputs $A$ is not scalable, then $A$ is not scalable. If $A$ is scalable, then the algorithm will output an $\eps$-scaling of $A$.
\end{thm}

It turns out that to test scalability, it suffices to take $\eps = 1/(n+1)$. More formally,

\begin{lem}[\cite{LSW}]\label{lem:eps=1/n-matrix} Suppose $A$ be an $n \times n$ non-negative matrix. If $A$ is row or column normalized and $\ds(A) \le 1/(n+1)$, then $A$ is scalable.
\end{lem}

Thus \cref{alg:matrix-scaling} along with \cref{thm:scalability_matrix} and \cref{thm:matrix-scaling} gives an alternate (albeit slower) algorithm to test if a bipartite graph has a perfect matching.\footnote{Note the iterates in \cref{alg:matrix-scaling} are row or column normalized and hence \cref{lem:eps=1/n-matrix} applies.}

\subsection{Operator scaling}\label{subsec:op_scaling}
 
The operator scaling problem was first introduced and studied by Gurvits \cite{gurvits2004classical}. We refer the reader to \cite{gurvits2004classical, garg2016deterministic, AGLOW18} for various motivations, connections and applications. The objects of study here are tuples of $n \times n$ complex matrices $A = (A_1,\ldots, A_m)$. The name operator scaling comes from the fact that these tuples define a map from positive definite matrices to themselves, by $T_A(X) = \sum_{i=1}^m A_i X A_i^{\dagger}$.\footnote{These maps are called completely positive maps/operators and are very natural from the point of view of quantum mechanics.}\footnote{$A_i^{\dagger}$ denotes the conjugate transpose of $A_i$.} But here we will restrict ourselves to the representation as tuple of matrices, for simplicity of exposition.

We start with a few definitions. First is the definition of scaling in this setting.

\begin{dfn}[Scaling of tuples]\label{dfn:op_scaling} Given a tuple of $n \times n$ complex matrices, $A = (A_1,\ldots, A_m)$, we say that $A' = (A'_1,\ldots, A'_m)$ is a scaling of $A$ if there exist invertible matrices $B,C$ s.t. $A' = B A C$ i.e. $A'_i = B A_i C$ for all $i$.
\end{dfn}

Next is the definition of doubly stochastic in this setting.

\begin{dfn}[Doubly stochastic tuples]\label{dfn:doubly_stochastic_op} A tuple of $n \times n$ complex matrices, $A = (A_1,\ldots, A_m)$, is said to be doubly stochastic if
$$
\sum_{i=1}^m A_i A_i^{\dagger} = \sum_{i=1}^m A_i^{\dagger} A_i = I_n
$$
\end{dfn}

As before, the operator scaling question is: given a tuple $A$, find a scaling which is doubly stochastic (if one exists). Again an approximate version is more natural. Towards that, we have the following definition quantifying how close a tuple is to being doubly stochastic.\footnote{We apologize for the overload of notation. Some of it is deliberate to draw out the syntactic similarity between various scaling problems. As we will see later, there is a common thread that binds all these problems.}

\begin{dfn}\label{dfn:ds_op} Given a tuple of $n \times n$ complex matrices, $A = (A_1,\ldots, A_m)$, define
$$
\ds(A) = \Norm{\sum_{i=1}^m A_i A_i^{\dagger} - I_n}_F^2 + \Norm{\sum_{i=1}^m A_i^{\dagger} A_i - I_n}_F^2
$$
Here $\norm{\cdot}_F$ denotes the Frobenius norm.
\end{dfn}

The goal in the current version of $\eps$-scaling problem is to find a scaling $A'$ of $A$ s.t. $\ds(A') \le \eps$ (if one exists).

\begin{dfn}[Scalability]\label{dfn:scalability_op} A tuple of $n \times n$ complex matrices, $A = (A_1,\ldots, A_m)$ is {\em scalable} if for all $\eps > 0$, there exists a scaling $A'$ of $A$ s.t. $\ds(A') \le \eps$.
\end{dfn}

We ask the same questions as before. When is a tuple scalable? If it is scalable, can we find an $\eps$-scaling efficiently? There is a deep theory underlying the answer to the first question, and to unveil it we will need another definition.\footnote{Notice the similarity with the definition of dimension expanders (see \cite{FG14} and references therein).}

\begin{dfn}[Dimension non-decreasing tuples]\label{dfn:dim_non-increasing} We say that a tuple of of $n \times n$ complex matrices, $A = (A_1,\ldots, A_m)$ is dimension non-decreasing if for all subspaces $V \subseteq \CC^n$, $\dim\left( \sum_{i=1}^m A_i(V)\right) \ge \dim(V)$. Here $A_i(V)$ denotes the subspace $\{A_i v : v \in V\}$ and $V + W$ denotes the subspace $\textnormal{span}\{v+w : v \in V, w \in W\}$. 
\end{dfn}

The following theorem gives a very pleasing answer to the first question.

\begin{thm}[\cite{gurvits2004classical}]\label{thm:scalability_operator} A tuple of $n \times n$ complex matrices, $A = (A_1,\ldots, A_m)$ is scalable iff $A$ is dimension non-decreasing.
\end{thm}

What about the second question? Gurvits \cite{gurvits2004classical} suggested a natural algorithm similar to that 
of Sinkhorn, although he could not analyze it in all cases. The full analysis, stated in the following theorem, 
was proved in \cite{garg2016deterministic}.

\begin{Algorithm}[th!]
\textbf{Input}:
A tuple of $n \times n$ matrices, $A = (A_1,\ldots, A_m)$ with entries in $\QQ$ having bit complexity at most $b$
and a distance parameter $\eps > 0$.
\\[.1ex]

\textbf{Output:}
Either the algorithm correctly identifies that $A$ is not scalable, or it outputs invertible matrices $B, C$ s.t. $\ds(B A C) \le \eps$.\footnotemark
\\[.1ex]

\textbf{Algorithm:}\vspace{-.2cm}

\begin{enumerate}
\item Initialize $B = C = I$.
\item If $\sum_{i=1}^m A_i A_i^{\dagger}$ or $\sum_{i=1}^m A_i^{\dagger} A_i$ is not invertible, output not scalable and halt.
\item Iterate for $T$ iterations:
\begin{itemize}
\item If $\ds(B A C) \le \eps$, output $B, C$ and halt. 
\subitem Else, if $\Norm{\sum_{i=1}^m A_i A_i^{\dagger} - I_n}_F^2 > \eps/2$, left normalize i.e. 
$$B \leftarrow \left( \sum_{i=1}^m A_i A_i^{\dagger}\right)^{-1/2} B$$ 
\subitem Otherwise right normalize (which can be defined analogously).
\end{itemize}
\item Output that $A$ is not scalable.
\end{enumerate}

\caption{Gurvits' algorithm}\label{alg:operator-scaling}
\end{Algorithm}
\footnotetext{Here $B A C = (B A_1 C, \ldots, B A_m C)$.}

\begin{thm}[\cite{garg2016deterministic}]\label{thm:alg-op} \cref{alg:operator-scaling} with $T = O(n(b+\log(n))/\eps)$ iterations works correctly. That is if the algorithm outputs $A$ is not scalable, then $A$ is not scalable. If $A$ is scalable, then the algorithm will output an $\eps$-scaling of $A$.
\end{thm}

Similar to the matrix scaling setting, to test scalability, it suffices to take $\eps = 1/(n+1)$. More formally,

\begin{lem}[\cite{gurvits2004classical}] Suppose $A = (A_1,\ldots, A_m)$ is a tuple of $n \times n$ complex matrices. If $\sum_{i=1}^m A_i A_i^{\dagger} = I_n$ or $\sum_{i=1}^m A_i^{\dagger} A_i = I_n$, and $\ds(A) \le 1/(n+1)$, then $A$ is scalable.
\end{lem}

Hence \cref{alg:operator-scaling} along with \cref{thm:scalability_operator} and \cref{thm:alg-op} gives a polynomial time algorithm to test if a tuple is dimension non-decreasing.

\begin{thm}[\cite{garg2016deterministic}] There is a polynomial time algorithm to test if a tuple $A = (A_1,\ldots, A_m)$ of $n \times n$ complex matrices is dimension non-decreasing.
\end{thm}

This was the first polynomial time algorithm for the operator scaling problem and as we will later see 
has applications in derandomization. Soon after, \cite{ivanyos2017constructive} (also 
see \cite{ivanyos2017noncommutative, derksen2015}) designed an algebraic algorithm for this problem which also 
works over finite fields. Their algorithm is an algebraic analogue of the augmenting paths algorithm for matching!

\subsection{Tensor scaling}\label{subsec:tensor_scaling}

In this section, we will discuss a scaling problem which is a generalization of operator scaling and was studied 
in~\cite{burgisser2017alternating}. The objects of study here are tuples of tensors. Let us denote the space 
of tensors $\CC^{n_1} \otimes \cdots \otimes \CC^{n_d}$ by $\Ten(n_1,\ldots, n_d)$. Then we will use the notation 
$A = (A_1,\ldots, A_m)$ to denote tuples of tensors where each $A_i \in \Ten(n_1,\ldots, n_d)$.

We start with the definition of scaling in this setting.\footnote{The scaling here looks very different from matrix scaling. One can also define a generalization of matrix scaling to tensors but we will not focus on that version in this survey (see \cite{FRANKLIN1989717}).}

\begin{dfn}[Tensor scaling of tuples]\label{dfn:tensor_scaling} Given a tuple of tensors (in $\Ten(n_1,\ldots, n_d)$), $A = (A_1,\ldots, A_m)$, we say that $A' = (A'_1,\ldots, A'_m)$ is a scaling of $A$ if there exist invertible matrices $g_1,\ldots, g_d$ s.t. $A'_i = (g_1 \otimes \cdots \otimes g_d) A_i$ for all $i$. We will use the notation $A' = (g_1, \ldots, g_d) \cdot A$ for this scaling action.
\end{dfn}

Before going to the definition of stochastic tuples in this setting, we need to define a certain notion of marginals.

\begin{dfn}[Marginals] Given a tuple of tensors, $A = (A_1,\ldots, A_m)$, identify it with $\Ten(m, n_1,\ldots, n_d)$. Then we will denote the marginals of $A$ by $\rho^A_1, \ldots, \rho^A_d$, where $\rho^A_i \in M_{n_i, n_i}(\CC)$ is a positive semidefinite matrix for all $i$. For each $i$, we can flatten $A$ to obtain $B_i \in M_{n_i, m \prod_{j \neq i} n_j}(\CC)$. Then $\rho^A_i = B_i B_i^{\dagger}$. These are uniquely characterized by the following property:
$$
\tr\left[(I_m \otimes I_{n_1} \otimes \cdots \otimes C_i \otimes \cdots I_{n_d}) A A^{\dagger} \right] = \tr\left[ C_i \rho^A_i\right]
$$
for all $C_i \in M_{n_i}(\CC)$ and for all $i \in [d]$.
\end{dfn}

\begin{rem}
The above notion of marginals is very natural from the point of view of quantum mechanics. If one views $A$ as representing a quantum state on $d+1$ systems indexed by $0,1,\ldots, d$, then $\rho^A_1, \ldots, \rho^A_d$ are the marginal states on systems $1,\ldots, d$ respectively.
\end{rem}

Now we are ready to define the notion of stochasticity in this setting.

\begin{dfn}[$d$-stochastic tuples] A tuple of tensors (in $\Ten(n_1,\ldots, n_d)$) $A = (A_1,\ldots, A_m)$ is said to be $d$-stochastic if for each $i$, $\rho^A_i = \frac{1}{n_i} I_{n_i}$ i.e. the marginals are all scalar multiples of identity matrices.
\end{dfn}

The normalization by $1/n_i$ is needed because $\tr\left[ \rho^A_i\right] = \norm{A}_2^2$ for all $i$. We will also need the following measure which quantifies how close a tuple is to being $d$-stochastic.

\begin{dfn}\label{dfn:ds_tensor} Given a tuple of tensors, $A = (A_1,\ldots, A_m)$, define
$$
\ds(A) = \sum_{i=1}^d \Norm{\rho^A_i - \frac{1}{n_i} I_{n_i}}_F^2
$$
\end{dfn}

Note that the definition above differs from \cref{dfn:ds_matrix,,dfn:ds_op} slightly in terms of a normalization factor. As before the $\eps$-scaling problem is to find a scaling $A'$ of $A$ s.t. $\ds(A') \le \eps$ (if one exists). Scalability is also defined similarly as before.

\begin{dfn}[Scalability]\label{dfn:scalability_tensor} We will say that a tuple of tensors, $A = (A_1,\ldots, A_m)$, is scalable if for all $\eps > 0$, there exists a scaling $A'$ of $A$ s.t. $\ds(A') \le \eps$.
\end{dfn}

The same questions arise. When is a tuple scalable? If it is scalable, can one find an $\eps$-scaling efficiently? The answer to the first question is given by remarkable and deep theorems of Hilbert and Mumford, and Kempf and Ness. To properly state this answer we need some more definitions.

\begin{dfn}[Deficiency]\label{dfn:deficient} We call a subset $S \subseteq [n_1] \times \cdots \times \cdots [n_d]$ deficient if there exist real numbers $(a_{i,j})_{i \in [d], j \in [n_i]}$ s.t. $\sum_{i = 1}^{d} a_{i, j_i} > 0$ for all $(j_1,\ldots, j_d) \in S$.
\end{dfn}

We encourage the reader to work out an alternate characterization of deficiency in the case $d = 2$ and $n_1 = n_2$. Hint: it is related to perfect matchings in bipartite graphs.

We will also use the following notation.
$$
\supp(A) = \{(j_1,\ldots, j_d) \in [n_1] \times \cdots \times [n_d]: \exists i \in [m] \: \text{s.t.} \: A_i(j_1,\ldots, j_d) \neq 0\}
$$

\begin{thm}[Hilbert-Mumford + Kempf-Ness \cite{Hil, Mum65, KN79}, see \cite{burgisser2017alternating}]\label{thm:scalability_tensor} A tuple of tensors $A = (A_1,\ldots, A_m)$ is scalable iff for every tuple of invertible matrices $(g_1,\ldots, g_d)$, $\supp((g_1,\ldots, g_d) \cdot A)$ is not deficient.
\end{thm}

We leave it as an exercise to verify that the above theorem is the same as \cref{thm:scalability_operator} in the case $d = 2$ and $n_1 = n_2$.

How to find an efficient scaling if one exists? It turns out that one can extend the alternating minimization kind of algorithms from the matrix and operator scaling settings to the tensor scaling setting as well.

\begin{Algorithm}[th!]
\textbf{Input}:
A tuple of tensors (in $\Ten(n_1,\ldots, n_d)$), $A = (A_1,\ldots, A_m)$ with entries in $\QQ$ having bit complexity at most $b$ and a distance parameter $\eps > 0$.
\\[.1ex]

\textbf{Output:}
Either the algorithm correctly identifies that $A$ is not scalable, or it outputs invertible matrices $(g_1,\ldots, g_d)$ s.t. $\ds((g_1,\ldots, g_d) \cdot A) \le \eps$.\footnotemark
\\[.1ex]

\textbf{Algorithm:}\vspace{-.2cm}

\begin{enumerate}
\item Initialize $g_i = I_{n_i}$.
\item If for some $i$, $\rho^A_i$ is not invertible, output not scalable and halt.
\item Iterate for $T$ iterations:
\begin{itemize}
\item If $\ds((g_1,\ldots, g_d) \cdot A) \le \eps$, output $(g_1,\ldots, g_d)$ and halt. If for some $i$, $\Norm{\rho^A_i - I_{n_i}}_F^2 > \eps/d$, normalize the $i^{\text{th}}$ coordinate i.e. $g_i \leftarrow \left(n_i \rho^A_i\right)^{-1/2} g_i$.
\end{itemize}
\item Output that $A$ is not scalable.
\end{enumerate}

\caption{Tensor scaling algorithm}\label{alg:tensor-scaling}
\end{Algorithm}
\footnotetext{Here $(g_1, \ldots, g_d) \cdot A = ((g_1 \otimes \cdots \otimes g_d) A_1, \ldots, (g_1 \otimes \cdots \otimes g_d) A_m)$.}

\cref{alg:tensor-scaling} was proposed in \cite{verstraete2003normal} without analysis. The following theorem regarding the analysis of the algorithm was proved in \cite{burgisser2017alternating}.

\begin{thm}[\cite{burgisser2017alternating}]\label{thm:alg-tensor} \cref{alg:operator-scaling} with $T = O(d (b+\log(m n_1 \cdots n_d))/\ell \eps)$ iterations works correctly $(\ell = \min_i n_i)$. That is if the algorithm outputs $A$ is not scalable, then $A$ is not scalable. If $A$ is scalable, then the algorithm will output an $\eps$-scaling of $A$.
\end{thm}

Unfortunately, unlike the matrix and operator scaling case, to test scalability, it is not sufficient to take $\eps$ which is polynomially small (see \cite{burgisser2017alternating} for a discussion). Hence we still do not have a polynomial time algorithm for testing scalability of tensors.

\section{Source of scaling}\label{sec:inv-theory}

Given the syntactic similarities between \cref{subsec:matrix_scaling,,subsec:op_scaling,,subsec:tensor_scaling}, it is natural to wonder if there is a general setting which captures all these scaling problems. In other words, where does scaling come from? It turns out that scaling arises in an algebraic setting and understanding the algebraic setting is crucial to a unified analysis of \cref{alg:matrix-scaling,,alg:operator-scaling,,alg:tensor-scaling}.

In \cref{subsec:inv_theory}, we introduce basic concepts in invariant theory, which provides crucial tools for 
the analysis of scaling algorithms. In \cref{subsec:git} we introduce basic concepts of geometric invariant 
theory, which elucidates the connection between invariant theory and scaling problems.

\subsection{Invariant theory: source of scaling}\label{subsec:inv_theory}


Invariant theory studies the linear actions of groups on vector spaces. We refer the reader to the excellent books \cite{derksen2015computational, sturmfels2008algorithms} for an extensive introduction to the area. We will only cover a few basics that we need for our purpose here. Invariant theory deals with linear actions of groups on vector spaces. For our purpose vector spaces will be over complex numbers ($\CC$) and the groups we will deal with will be extremely simple - special linear group, denoted by $\SL(n)$ ($n \times n$ matrices over $\CC$ with determinant $1$), direct products of special linear group as well as the diagonal subgroup of the special linear group, denoted by $\ST(n)$ (diagonal $n \times n$ matrices over $\CC$ with determinant $1$) and direct products. However the theory is quite general and generalizes to large class of groups.

Suppose we have a group $G$ which acts linearly on a vector space $V$.\footnote{That is the group action satisfies the following axioms: $g \cdot (c_1 v_1 + c_2 v_2) = c_1 g \cdot v_1 + c_2 g \cdot v_2$ for all $c_1, c_2 \in \CC$ and $v_1, v_2 \in V$, in addition to the properties of being a group action i.e. $g_1 \cdot (g_2 \cdot v) = (g_1 g_2) \cdot v$, $e \cdot v = v$ for all $g_1, g_2 \in G$, $v \in V$ and for $e$ being the identity element of the group. Usually one also requires that the action is algebraic.} Fundamental objects of study in invariant theory are the invariant polynomials which are just polynomial functions on $V$ left invariant by the action of the group $G$. Invariant polynomials form a ring and this ring is usually denoted by $\CC[V]^G$. More formally,
$$
\CC[V]^G = \{p \in \CC[V]: p(g \cdot v) = p(v) \: \forall \: g \in G, v \in V\}
$$

Let us consider a simple example. The group $G = \SL(n) \times \SL(n)$ acts on the vector space $V = \mat_n(\CC)$ \footnote{$\mat_n(\CC)$ denotes the space of $n \times n$ complex matrices} by left-right multiplication as follows: $(A,B) \cdot X = A X B^T$. $\Det(X)$ is an invariant polynomial for this action and it turns out it is the only one (prove it!). That is any invariant polynomial is just of the form $q(\Det(X))$, for a univariate polynomial $q$, or in other words, $\Det(X)$ generates the invariant ring. As an aside (this will not be so important for us), Hilbert \cite{Hil90, Hil} proved that the invariant ring is always finitely generated! \footnote{He proved it for the actions of general linear groups but his proof readily generalizes to a more general class of groups called reductive groups.} These papers proved several theorems which are the building blocks of modern algebra, like Nullstellansatz and finite basis theorem, as ``lemmas" enroute to proving the finite generation of invariant rings!

Some other fundamental objects of study in invariant theory are orbits and orbit-closures. The orbit of a vector $v \in V$, $\cO_G(v)$ is simply the set of all vector elements that $v$ can be transformed to by the group action. That is,
$$
\cO_G(v) = \{g \cdot v: g \in G\}
$$
An orbit-closure, $\clO_G(v)$ of a vector $v$ is obtained by simply including all the limit points of sequences of points in an orbit. That is,
$$
\clO_G(v) = \{w \in V: \exists g_1,\ldots, g_k,\ldots, \: \text{s.t.} \: \text{lim}_{k \rightarrow \infty} g_k \cdot v = w\}
$$
Many important problems in theoretical computer science are really questions about orbit-closures. To list a few,
\begin{enumerate}
\item The graph isomorphism problem is about checking if the orbit closures\footnote{Note that for the action of a finite group, the orbit of a point is the same as its orbit closure.} of two graphs (under the group action of permuting the vertices) are the same or not.
\item The $\mathcal{VP}$ vs $\mathcal{VNP}$ question (or more precisely a variant of it) can be phrased as testing if the (padded) permanent polynomial lies in the orbit-closure of the determinant (w.r.t. the action on the polynomials induced by the action of general linear group on the variables). This is the starting point of geometric complexity theory (GCT) \cite{Mulmuley_Sohoni:2002, Bugisser_GCT, Landsberg2015}.
\item The question of tensor rank lower bounds (more precisely border rank) can be phrased as asking if a padded version of the given tensor lies in the orbit-closure of the diagonal unit tensor (w.r.t. the natural action of products of general linear groups on the tensors). This approach also falls under the purview of geometric complexity theory \cite{burgisser2011geometric, burgisser2013explicit}.
\end{enumerate}

It turns out that a very simple concept in invariant theory captures the mysteries about the scaling problems in \cref{subsec:matrix_scaling,,subsec:op_scaling,,subsec:tensor_scaling}. This is the so called \emph{null cone} of a group action (on a vector space). The null cone has dual definitions in terms of the invariant polynomials as well as orbit-closures (in a very general setting, and in particular for the group actions we care about in this survey). This duality is quite important for the analysis of the scaling algorithms.

\begin{dfn}[Null cone]\label{dfn:nullcone} The null cone for a group $G$ acting on a vector space $V$, denoted by $\cN_G(V)$, is the zero set of all homogeneous invariant polynomials. That is,
$$
\cN_G(V) = \{v \in V: p(v) = 0 \: \forall \: \text{homogeneous} \: p \in \CC[V]^G\}
$$
\end{dfn}

It is a cone since $v \in \cN_G(V)$ implies that $c v \in \cN_G(V)$ for all $c \in \CC$. A theorem due to Hilbert \cite{Hil} and Mumford \cite{Mum65} \footnote{Not to be confused with Hilbert-Mumford criterion which we will come across later.} says that for a large class of group actions (which includes the group actions we will study), $v \in \cN_G(V)$ iff $0 \in \clO_G(v)$ (try to figure out the easy direction). This is a consequence of Hilbert's Nullstellensatz along with the fact that orbit-closures for certain group actions are algebraic varieties (or in other words Euclidean and Zariski closures match). If we look at the left-right multiplication example discussed above, the null cone is just the space of singular matrices since determinant generates the invariant ring. We leave it as an exercise to verify that the $0$ matrix lies in the orbit-closure of any singular matrix (under the left-right multiplication action of $\SL(n) \times \SL(n)$). 

We will now describe the connection between null cone and scaling problems. For this we will need to move on to the area of geometric invariant theory, which provides geometric and analytic tools to study problems in invariant theory, and also provides with an intriguing non-commutative extension of Farkas' lemma (or linear programming duality). As a teaser of things to come, the objects in \cref{subsec:matrix_scaling,,subsec:op_scaling,,subsec:tensor_scaling} are scalable iff they are not in the null cone of certain group actions!

\subsection{Geometric invariant theory: non-commutative duality}\label{subsec:git}

In this section, we will give a brief overview of the geometric invariant theoretic approach to studying the null cone problem. This will also fit in nicely with the computational aspects of the null cone. \cref{subsubsec:HM} describes the Hilbert-Mumford criterion which is really answering the question: how does one prove if some vector is in the null cone. \cref{subsubsec:KN} describes Kempf-Ness which answers the question: how does one prove if some vector is not in the null cone. \cref{subsubsec:Farkas} studies the Hilbert-Mumford and the Kempf-Ness criterion for certain commutative group actions and explains why these generalize Farkas' lemma. \cref{subsec:HM_KN_scaling} explains the connection between geometric invariant theory and scaling problems.

\subsubsection{Hilbert-Mumford criterion}\label{subsubsec:HM}

Fix the action of a group $G$ on a vector space $V$. How does one prove to someone that a vector $v$ is in the null cone? We know that $v$ is in the null cone iff $0 \in \clO_G(v)$ i.e. there is a sequence of group elements $g_1,\ldots, g_k,\ldots$ s.t. $\text{lim}_{k \rightarrow \infty} g_k \cdot v = 0$. So this sequence of group elements is a witness to $v$ being in the null cone. Is their a more succinct witness? After all, how do we even describe an infinite sequence of group elements? The Hilbert-Mumford criterion says that there does exist a more succinct witness (again we won't go into the technical conditions the group $G$ needs to satisfy but just say that they will be satisfied for the groups we will consider).

\begin{thm}[Hilbert-Mumford criterion \cite{Hil,Mum65}]\label{thm:HM} $v \in \cN_G(V)$ iff there is a one-parameter subgroup $\lambda$ of $G$ s.t. $\textnormal{lim}_{t \rightarrow 0} \lambda(t) \cdot v = 0$.
\end{thm}

What this means is that instead of looking at all sequences of group elements, one only needs to restrict our attention to those sequences of group elements which can be succinctly described by one-parameter subgroups. What are one-parameter subgroups? These are just algebraic group homomorphisms (i.e. an algebraic map which is also a group homomorphism) $\lambda : \CC^* \rightarrow G$. Let us look at several examples (we encourage the reader to prove these statements). 

\begin{enumerate}
\item For the group $G = \CC^*$ (the multiplicative group of non-zero complex numbers), all one parameter subgroups are of the form $\lambda(t) = t^a$ for some $a \in \ZZ$.
\item For the group $G = \left(\CC^{*}\right)^n$ (direct product of $n$ copies of $\CC^*$), all one parameter subgroups are of the form $\lambda(t) = \left(t^{a_1},\ldots, t^{a_n}\right)$ for some $(a_1,\ldots, a_n) \in \ZZ^n$.
\item For the group $G = \ST(n)$ (diagonal $n \times n$ matrices with determinant $1$), all one parameter subgroups are of the form $\lambda(t) = \left(t^{a_1},\ldots, t^{a_n}\right)$ for some $(a_1,\ldots, a_n) \in \ZZ^n$ satisfying $\sum_{i=1}^n a_i = 0$.
\item For the group $G = \ST(n) \times \ST(n)$, all one parameter subgroups are of the form 
$$
\lambda(t) = \left(\left(t^{a_1},\ldots, t^{a_n}\right), \left(t^{b_1},\ldots, t^{b_n}\right)\right)
$$ 
for some $(a_1,\ldots, a_n), (b_1,\ldots, b_n) \in \ZZ^n$ satisfying $\sum_{i=1}^n a_i = \sum_{i=1}^n b_i = 0$.
\item For the group $G = \GL(n)$ ($n \times n$ invertible matrices), all one parameter subgroups are of the form $\lambda(t) = S \: \text{diag}\left(t^{a_1},\ldots, t^{a_n} \right) S^{-1}$ for some $S \in \GL(n)$ and some $(a_1,\ldots, a_n) \in \ZZ^n$. Here $\text{diag}\left( t_1,\ldots, t_n\right)$ represents a diagonal matrix with $(t_1,\ldots, t_n)$ on the diagonal.
\item For the group $G = \SL(n)$ ($n \times n$ invertible matrices with determinant $1$), all one parameter subgroups are of the form $\lambda(t) = S \: \text{diag}\left(t^{a_1},\ldots, t^{a_n} \right) S^{-1}$ for some $S \in \SL(n)$ and some $(a_1,\ldots, a_n) \in \ZZ^n$ satisfying $\sum_{i=1}^n a_i = 0$.
\item For the group $G = \SL(n_1) \times \cdots \SL(n_d)$, all one parameter subgroups are of the form 
$$
\lambda(t) = \left(S_1 \: \text{diag}\left(t^{a_{1,1}},\ldots, t^{a_{1,n_1}} \right) S_1^{-1} , \ldots, S_d \: \text{diag}\left(t^{a_{d,1}},\ldots, t^{a_{d,n_d}} \right) S_d^{-1}\right)
$$ 
for some $S_i \in \SL(n_i)$ and some integer $a_{i,j}$'s satisfying $\sum_{j=1}^{n_i} a_{i,j} = 0$ for all $i \in [d]$.
\end{enumerate}

Let us return to the example of the left-right multiplication action of $G = \SL(n) \times \SL(n)$ on $V = \mat_n(\CC)$. Recall that $(A,B)$ sends $M$ to $A M B^T$ and $M$ is in the null cone iff it is singular. If $M$ is singular, what is a one-parameter subgroup driving it to the zero matrix? Since $M$ is singular, there exists an invertible $S$ (which can be taken to have determinant $1$) s.t. $S^{-1} M$ has the last row all zeroes. Then the one-parameter subgroup 
$$
\lambda(t) = \left( S \: \text{diag} \left( t,t,\ldots,t, t^{-(n-1)}\right) S^{-1}, I_n\right)
$$
sends $M$ to the zero matrix. Later we will see more examples corresponding to each of the scaling problems. 

Having understood how to prove if a given vector is in the null cone, we move on to study how to prove that a 
given vector is {\em not} in the null cone.

\subsubsection{Kempf-Ness theorem}\label{subsubsec:KN}

Fix the action of a group $G$ on a vector space $V$. How does one prove that a vector $v$ is not in the null cone? We know that a vector $v \notin \cN_G(V)$ iff there is a homogeneous invariant polynomial $p$ s.t. $p(v) \neq 0$. Such a $p$ can serve as a witness that $v \notin \cN_G(V)$. However, these polynomials typically have exponentially large degree (see \cite{derksen2001polynomial}) and may not have any efficient description. An alternative witness is given by the Kempf-Ness theorem \cite{KN79}.

To state the Kempf-Ness theorem, we need to (informally) define something called a moment map, which relies on the following function,
$$
f_v(g) = \norm{g \cdot v}_2^2
$$
This function defines the following optimization problem,
\begin{align}
\cN(v) = \text{inf}_{g \in G} f_v(g) \label{eqn:capacity}
\end{align}
Note that $v \notin \cN_G(V)$ iff $\cN(v) > 0$. Now the moment map at $v$, denoted by $\mu_G(v)$ is simply the gradient of the function $f_v$ ``along the group action'' at $g = e$ (the identity element of the group $G$).\footnote{There are minor differences between this definition and how moment map is usually defined.} We will not go into the specifics of the space in which $\mu_G(v)$ lives but instead do the moment map calculation for several examples. First let us state the Kempf-Ness theorem.

\begin{thm}[Kempf-Ness \cite{KN79}]\label{thm:KN} $v \notin \cN_G(V)$ iff there is a non-zero $w \in \clO_G(v)$ s.t. $\mu_G(w) = 0$.
\end{thm}

If $v \notin \cN_G(V)$, then there exists a non-zero $w \in \clO_G(v)$ which is of minimal norm and hence $\mu_G(w) = 0$.  So this is the easy direction. The amazing part about the Kempf-Ness theorem is that any local minima becomes a global minima i.e. if $\mu_G(w) = 0$ for some non-zero $w \in \clO_G(v)$, then $v \notin \cN_G(V)$, even though $\mu_G(w) = 0$ only guarantees that one cannot decrease the norm of $w$ by actions of group elements close to identity (that is, ``local'' group actions). This smells of some kind of convexity and indeed, the function $f_v(g)$ is geodesically convex (i.e. convex w.r.t. some appropriate metric on the group). We will not delve more into geodesic convexity or moment maps in this survey but refer the interested reader to \cite{ness1984stratification, Woodward11, Hechman-Hochs12, georgoulas2013moment}.

Let us return to the example of the left-right multiplication action of $G = \SL(n) \times \SL(n)$ on $V = \mat_n(\CC)$. Recall that $(A,B)$ sends $M$ to $A M B^T$ and $M$ is in the null cone iff it is singular. What is the moment map in this case? $\mu_G(M) = (P_1, P_2)$, where $P_1, P_2$ are $n \times n$ traceless matrices s.t.
\begin{align*}
\tr\left[P_1 Q_1\right] + \tr\left[P_2 Q_2\right] &= \frac{d}{ds} \Norm{\exp\left(s Q_1\right) M \exp\left(s Q_2^T\right)}_F^2 \bigg|_{s=0} \\
&= 2 \tr[M M^{\dagger} Q_1] + 2 \tr\left[\left(M^{\dagger} M \right)^T Q_2\right]
\end{align*}
for all Hermitian traceless matrices $Q_1, Q_2$. \footnote{It suffices to focus on Hermitian matrices since for skew-Hermitian matrices, their exponential is unitary and hence cannot change the norm. Also note that $Q \rightarrow \tr\left[P Q\right]$ is an $\mathbb{R}$-linear map over the space of Hermitian matrices $Q$, when $P$ is Hermitian. A crucial point here which we have glossed over, but nonetheless the reader should verify, is that if $\ell : M_n(\CC) \rightarrow \RR$ is an $\RR$-linear map, then there exists a Hermitian $P$ s.t. for all Hermitian $Q$, $\ell(Q) = \tr[P Q]$.} Here $M^{\dagger}$ denotes the conjugate transpose of the matrix $M$. Thus
$$
P_1/2 = MM^{\dagger} - \frac{\norm{M}_F^2}{n} I_n \: \text{and} \: P_2/2 = \left( M^{\dagger} M \right)^T - \frac{\norm{M}_F^2}{n} I_n
$$
Hence $\mu_G(M) = 0$ is the same as saying $M$ is a scalar multiple of a unitary matrix. It is not hard to see that any non-singular $M$ can be brought to such a form by the left-right multiplication action.

In the next section, we will see what Hilbert-Mumford and Kempf-Ness theorem look like for actions of $\T(n) = \left(\CC^* \right)^n$. Readers wanting to get to the setting of scaling problems could skip the next section.

\subsubsection{Commutative group actions: Farkas' lemma}\label{subsubsec:Farkas}

In this section, we play around with the Hilbert-Mumford criterion and Kempf-Ness theorem and see what it gives for actions of the group $G = \T(n) = \left(\CC^* \right)^n$.

Fix vectors $\omega^{(1)},\ldots, \omega^{(m)} \in \ZZ^n$. Then $G$ acts on $V = \CC^m$ as follows: $(t_1,\ldots, t_n)$ sends the $i^\text{th}$ basis vector $e_j$ to $\prod_{i=1}^n t_i^{\omega^{(j)}_i} e_j$. That is $e_j$ is an eigenvector of the action of $(t_1,\ldots, t_n)$ with eigenvalue $\prod_{i=1}^n t_i^{\omega^{(j)}_i}$. We urge the reader to prove that all actions of $G$ look essentially like this.

What is the null cone for this action? Let us apply Hilbert-Mumford criterion. Recall that all the one-parameter subgroups of $G$ look like $\lambda(t) = \left( t^{a_1},\ldots, t^{a_n}\right)$ for some $(a_1,\ldots, a_n) \in \ZZ^n$. Now fix $v \in V$, where $v = \sum_{j=1}^m v_j e_j$, with $v_j \in \CC$, and denote by $\text{supp}(v)$, the support of $v$ i.e.
$$
\text{supp}(v) = \{j \in [m]: v_j \neq 0\}
$$
Then the Hilbert-Mumford criterion (\cref{thm:HM}) tells us that $v \in \cN_G(V)$ iff there is a one-parameter subgroup $\lambda$ that drives $v$ to zero. That is, there exists $ a = (a_1,\ldots, a_n) \in \ZZ^n$ s.t.
\begin{align*}
\text{lim}_{t \rightarrow 0} \prod_{i=1}^n t^{a_i \omega^{(j)}_i} = \text{lim}_{t \rightarrow 0} t^{\langle a, \omega^{(j)} \rangle} = 0
\end{align*}
for all $j \in \text{supp}(v)$. Equivalently, we have: 

\begin{prp}\label{prp:HM_commutative}
$v \in \cN_G(V)$ iff there exists $a \in \ZZ^n$ s.t. $\langle a, \omega^{(j)} \rangle > 0$ for all $j \in \text{supp}(v)$.
\end{prp}

Now let us see what the Kempf-Ness theorem says in this setting. By computing the moment map $\mu_G(v) \in \RR^n$ we see that it satisfies the following,
\begin{align*}
\langle \mu_G(v), b \rangle &= \frac{d}{ds} \Norm{\left( \exp(sb_1),\ldots, \exp(sb_n)\right) \cdot v}_2^2 \bigg|_{s=0} \\
&= 2 \sum_{j=1}^m |v_j|^2 \langle \omega^{(j)}, b \rangle 
\end{align*}
for all $b \in \RR^n$.\footnote{Again as before, it suffices to look at $\RR^n$, since the imaginary part $i \RR^n$ (the exponential of it) does not change the norm.} Hence $\mu_G(v) = 2 \sum_{j=1}^m |v_j|^2 \omega^{(j)} = 2 \sum_{j \in \text{supp}(v)} |v_j|^2 \omega^{(j)}$. Now the Kempf-Ness theorem (\cref{thm:KN}) says that $v \notin \cN_G(V)$ iff there exists non-zero $w \in \clO_G(v)$ s.t. $\mu_G(w) = 0$. Note that if there exists non-zero $w \in \clO_G(v)$ s.t. $\mu_G(w) = 0$, then $0 \in \text{conv} \left( \left(\omega^{(j)}\right)_{j \in \text{supp}(v)}\right)$. So this matches with the conclusions of the Farkas' lemma which says that there exists $a \in \ZZ^n$ s.t. $\langle a, \omega^{(j)} \rangle > 0$ for all $j \in \text{supp}(v)$ iff $0 \notin \text{conv} \left( \left(\omega^{(j)}\right)_{j \in \text{supp}(v)}\right)$. The first part of the Farkas' lemma matches the case $v \in \cN_G(V)$ via the Hilbert-Mumford criterion and the second part matches the case $v \notin \cN_G(V)$ via the Kempf-Ness theorem!



\subsection{Hilbert-Mumford, Kempf-Ness and scaling}\label{subsec:HM_KN_scaling}

In this section, we delve into the connection between geometric invariant theory and various scaling problems. \cref{subsubsec:matrix_scaling,,subsubsec:op_scaling,,subsubsec:tensor_scaling} consider the consequences of Hilbert-Mumford and Kemp-Ness theorems for the matrix, operator and tensor scaling problems, respectively.

\subsubsection{Matrix scaling}\label{subsubsec:matrix_scaling}

We elucidate here the connection between geometric invariant theory and matrix scaling. For the connection to invariant theory, we need a group action on a vector space. Given that the objects of study are $n \times n$ non-negative real matrices, it is natural to guess the vector space would be $V = \mat_n(\CC)$ (given that we only discussed invariant theory with the base field being $\CC$). But what is the group action? The group action is also almost given away by the definition of scaling. The first guess might be that the group is $G = \T(n) \times \T(n)$ and it acts on $V = \mat_n(\CC)$ as follows,
$$
\left((t_1,\ldots,t_n), (s_1,\ldots, s_n) \right) \cdot M = \text{diag}(t_1,\ldots,t_n) \: M \: \text{diag}(s_1,\ldots,s_n)
$$
But it turns out that the null cone for this action is the whole of $V$ (verify this). But the above guess comes pretty close and the right thing is obtained by looking at an appropriate normalization. It turns out that the group $G$ would be $\ST(n) \times \ST(n)$ and it acts on $V$ by the same action as above (it won't be immediately clear why imposing a determinant $1$ constraint on the group elements is the right thing to do).

Now let us see what the Hilbert-Mumford criterion (\cref{thm:HM}) and Kempf-Ness theorem (\cref{thm:KN}) say about this group action.

Recall from \cref{subsubsec:HM} that all the one-parameter subgroups of $G$ look like
$$
\lambda(t) = \left(\left(t^{a_1},\ldots,t^{a_1}\right), \left(t^{b_1},\ldots, t^{b_n}\right) \right)
$$
for some integers $a_1,\ldots, a_n, b_1,\ldots, b_n$ satisfying $\sum_{i=1}^n a_i = \sum_{i=1}^n b_i = 0$. Let us denote by $\text{supp}(M)$, the support of $M$ i.e.
$$
\text{supp}(M) = \{(i,j) \in [n] \times [n]: M_{i,j} \neq 0\}
$$
Then the Hilbert-Mumford criterion says that $M$ is in the null cone iff there exists a one-parameter subgroup $\lambda$ as above s.t.
$$
\text{lim}_{t \rightarrow 0} \lambda(t) \cdot M = 0
$$
Equivalently,

\begin{cor}\label{prp:HM_matrix_scaling} $M \in \cN_G(V)$ iff there exist integers $a_1,\ldots, a_n, b_1,\ldots, b_n$ satisfying $\sum_{i=1}^n a_i = \sum_{i=1}^n b_i = 0$ s.t. $a_i + b_j > 0$ for all $(i,j) \in \text{supp}(M)$.
\end{cor}

We encourage the reader to prove that the above proposition implies that $M \in \cN_G(V)$ iff the bipartite graph defined by $\text{supp}(M)$ has no perfect matching.

Now let us apply the Kempf-Ness theorem. First let us calculate the moment map. $\mu_G(M) = (p,q)$, where $p,q \in \RR^n$ and $\sum_{i=1}^n p_i = \sum_{j=1}^n q_j = 0$ and it satisfies the following,
\begin{align*}
\langle p, d \rangle + \langle q, e \rangle &= \frac{d}{ds} \Norm{\left(\left( \exp(sd_1),\ldots, \exp(sd_n)\right), \left( \exp(se_1),\ldots, \exp(se_n)\right)\right) \cdot M}_F^2 \bigg|_{s=0} \\
&= 2 \sum_{i,j} |M|_{i,j}^2 (d_i + e_j) \\
&= 2 \langle r_M, d \rangle + 2 \langle c_M, e \rangle
\end{align*}
for all $d,e \in \RR^n$ satisfying $\sum_{i=1}^n d_i = \sum_{j=1}^n e_j = 0$. Here $r_M$ and $c_M$ are the vectors of row and column sums of the matrix $\left( |M_{i,j}|^2\right)_{i \in [n], j \in [n]}$, respectively. Thus $p = r_M - \text{avg}_M \vec{1}$ and $q = c_M - \text{avg}_M \vec{1}$, where 
$$
\text{avg}_M = \sum_{i=1}^n r_M(i)/n = \sum_{j=1}^n c_M(j)/n
$$
and $\vec{1}$ is the all $1$'s vector. Now the Kempf-Ness theorem says that $M \notin \cN_G(V)$ iff there exists a non-zero $N \in \clO_G(M)$ s.t. $\mu_G(N) = 0$. Equivalently, 

\begin{cor}\label{prp:KN_matrix_scaling} $M \notin \cN_G(V)$ iff the non-negative real matrix $A_M$, given by $A_M(i,j) = |M_{i,j}|^2$, is scalable.
\end{cor}

\cref{prp:HM_matrix_scaling,,prp:KN_matrix_scaling} together yield a proof of \cref{thm:scalability_matrix}.

\subsubsection{Operator scaling}\label{subsubsec:op_scaling}

For the operator scaling problem, the vector space is clear, $V = \mat_n(\CC)^m$, i.e. $m$ copies of $\mat_n(\CC)$. The group action is also clear (except for the normalization to determinant $1$). $G = \SL(n) \times \SL(n)$ and it acts on $V$ as follows,
$$
(B,C) \cdot (A_1,\ldots, A_m) = (B A_1 C^T,\ldots, B A_m C^T)
$$
This action is sometimes called the left-right action. We leave the details of the Hilbert-Mumford criterion and Kempf-Ness theorem to the reader and only say that they yield the following corollaries which together imply \cref{thm:scalability_operator}.

\begin{cor}[Hilbert-Mumford for left-right action]
$A = (A_1,\ldots, A_m) \in \cN_G(V)$ iff $A$ is dimension non-decreasing (\cref{dfn:dim_non-increasing}).
\end{cor}

\begin{cor}[Kempf-Ness for left-right action]\label{cor:KN_left-right}
$A = (A_1,\ldots, A_m) \notin \cN_G(V)$ iff $A$ is scalable.
\end{cor}

\subsubsection{Tensor scaling}\label{subsubsec:tensor_scaling}

For the tensor scaling problem, the vector space is $V = \Ten(n_1,\ldots, n_d)^m$. The group is $G = \SL(n_1) \times \cdots \times \SL(n_d)$ which acts on $V$ as follows,
$$
(g_1,\ldots, g_d) \cdot (A_1,\ldots, A_m) = \left( \left(g_1 \otimes \cdots \otimes g_d \right) A_1,\ldots, \left(g_1 \otimes \cdots \otimes g_d \right) A_m\right)
$$
Again we will leave the details of the Hilbert-Mumford criterion and Kempf-Ness theorem to the reader and only say that they yield the following corollaries, which together imply \cref{thm:scalability_tensor}. 

\begin{cor}[Hilbert-Mumford for tensor action]
$A = (A_1,\ldots, A_m) \in \cN_G(V)$ iff there is a tuple of invertible matrices (of appropriate sizes) $(g_1,\ldots, g_d)$ s.t. $\text{supp} \left( (g_1,\ldots, g_d) \cdot A\right)$ is deficient (\cref{dfn:deficient}).
\end{cor}

\begin{cor}[Kempf-Ness for tensor action]\label{cor:KN_tensor}
$A = (A_1,\ldots, A_m) \notin \cN_G(V)$ iff $A$ is scalable.
\end{cor}

\section{Analysis of scaling algorithms}\label{sec:analysis}

In this section, we provide a unified analysis of the scaling algorithms described in \cref{subsec:matrix_scaling,,subsec:op_scaling,,subsec:tensor_scaling}. We will first design a common template and analysis for \cref{alg:matrix-scaling,,alg:operator-scaling,,alg:tensor-scaling} and then look at each case separately to fill in the details that need to be done differently. Most of the analysis will be common and the only difference will be the choice of a potential function (although the source of all the potential functions will be invariant theory). \cref{alg:template-scaling} contains a common template for all the three scaling algorithms.

\begin{Algorithm}[!htbp]
\textbf{Input}:
Object $A$ which has norm $\norm{A} = 1$, with all entries having bit complexity at most $b$ and distance parameter
$\eps > 0$. This means the following in various settings,
\begin{itemize}
\itemsep0em
\item {\bfseries Matrix scaling}: $A$ is a non-negative rational $n \times n$ matrix. $\norm{A} = \sum_{i,j} A_{i,j}$.
\item {\bfseries Operator scaling}: $A$ is a tuple of $n \times n$ matrices, $A = (A_1,\ldots, A_m)$, with entries in $\QQ$. $\norm{A} = \left(\sum_{i=1}^m \norm{A_i}_F^2\right)^{1/2}$.
\item {\bfseries Tensor scaling}: $A$ is a tuple of tensors (in $\Ten(n_1,\ldots, n_d) = \CC^{n_1} \otimes \cdots \otimes \CC^{n_d}$), $A = (A_1,\ldots, A_m)$, with entries in $\QQ$. $\norm{A} = \left(\sum_{i=1}^m \norm{A_i}_2^2\right)^{1/2}$.
\end{itemize}

\textbf{Output:}
Either the algorithm correctly identifies that $A$ is not scalable, or it outputs a scaling $A'$ of $A$ s.t. $\widetilde{\ds}(A') \le \eps$. The measure $\widetilde{\ds}$ is the same as $\ds$ for tensor scaling (\cref{dfn:ds_tensor}) while for matrix and operator scaling, there is a minor variation as explained below,\footnotemark
\begin{itemize}
\itemsep0em
\item {\bfseries Matrix scaling}: $\widetilde{\ds}(A) = \sum_{i=1}^n (r_i - 1/n)^2 + \sum_{j=1}^n (c_j - 1/n)^2$ (cf. \cref{dfn:ds_matrix}).
\item {\bfseries Operator scaling}: $\widetilde{\ds}(A) = \Norm{\sum_{i=1}^m A_i A_i^{\dagger} - I_n/n}_F^2 + \Norm{\sum_{i=1}^m A_i^{\dagger} A_i - I_n/n}_F^2$ (cf. \cref{dfn:ds_op}).
\end{itemize}

\textbf{Algorithm:}\vspace{-.4cm}

\begin{enumerate}
\item Check if some trivial conditions hold. If these do not hold, then output not scalable and halt.
\item There is a group $\widetilde{G}$ and a subset of the group $H \subseteq \widetilde{G}$. In essence $H$ corresponds to the normalization steps in the scaling algorithms from Section~\ref{sec:scaling_algs}. We will iteratively act on the current object by elements in $H$. $H$ will always be the subset corresponding to tuples of positive definite matrices s.t. at most one matrix in the tuple is not identity. For matrix scaling, $\widetilde{G} = \T(n) \times \T(n)$, which acts as in \cref{dfn:scaling_matrix}. For operator scaling, $\widetilde{G} = \GL(n) \times \GL(n)$, which acts as in \cref{dfn:op_scaling}. For tensor scaling, $\widetilde{G} = \GL(n_1) \times \cdots \times \GL(n_d)$, which acts as in \cref{dfn:tensor_scaling}.
\item Let $A^{(0)} = A$. For $T$ iterations, $t = 0$ to $T-1$: 
\begin{itemize}
\item If $\widetilde{\ds}\left(A^{(t)}\right) \le \eps$, then output $A^{(t)}$ and halt. Otherwise $A^{(t+1)} =h^{(t)} \cdot A^{(t)}$. $h^{(t)}$ is chosen according to some rule but all we will need for the analysis is that $\norm{A^{(t)}} = 1$ is preserved throughout and as a consequence the non-identity element in the tuple of matrices $h^{(t)}$, denoted by $\widehat{h^{(t)}}$, satisfies $\tr\left[\left(\widehat{h^{(t)}} \right)^{-k} \right] = n'$ ($k=1$ for matrix scaling and $k=2$ for operator and tensor scaling; we ask the reader to verify this in all the three cases). Also as a consequence of $\widetilde{\ds}\left(A^{(t)}\right) > \eps$, $\tr\left[ \left(\left(\widehat{h^{(t)}}\right)^{-k} - I_{n'} \right)^2\right] \ge \eps'$. Here $n'$ is the dimension of $h^{(t)}$ and $\eps' = (n')^2\eps/2$ for matrix and operator scaling, and $\eps' = (n')^2 \eps/d$ for tensor scaling.
\end{itemize}
\item Output that $A$ is not scalable.
\end{enumerate}

\caption{Common template for \cref{alg:matrix-scaling,,alg:operator-scaling,,alg:tensor-scaling} (has a different normalization).}\label{alg:template-scaling}
\end{Algorithm}
\footnotetext{This is to reconcile the minor differences in the definition of $\ds$ for various measures.}

Let us now turn to performing an analysis of \cref{alg:template-scaling}. We need a potential function and the source of potential functions will be invariant theory. Recall from \cref{subsec:HM_KN_scaling} that $A$ is a scalable iff $A \notin \cN_G(V)$ (\cref{prp:KN_matrix_scaling,,cor:KN_left-right,,cor:KN_tensor}).\footnote{\cref{prp:KN_matrix_scaling} says something slightly different but the variant we state here is true as well.} Here $G$ is a subgroup of $\widetilde{G}$, and $G = \ST(n) \times \ST(n)$, $\SL(n) \times \SL(n)$ or $\SL(n_1) \times \cdots \times \SL(n_d)$ for matrix, operator or tensor scaling, respectively. We also know from \cref{dfn:nullcone} that $A \notin \cN_G(V)$ iff there exists an $\ell \in \NN$ and a homogeneous polynomial of degree $\ell$ that is invariant under the action of $G$ s.t. $P(A) \neq 0$. Suppose there exists such a $P$ that has integer coefficients and satisfies
\begin{align}
|P(A)| \le U^\ell \norm{A}^\ell \label{eqn:ankit10}
\end{align}

Then we we will prove the following theorem regarding the analysis of \cref{alg:template-scaling}.

\begin{thm}[Unified analysis of scaling algorithms]\label{thm:unified} If $A$ is scalable, then running \cref{alg:template-scaling} for $T = O((\log(U) + b)/\eps'')$ iterations suffices to output a scaling $A'$ s.t. $\widetilde{\ds}(A') \le \eps$. Here $\eps'' = n\eps$ for matrix and operator scaling, and $\eps'' = \left(\min_i n_i\right) \eps/d$ for tensor scaling.
\end{thm}

The analysis will be a three step analysis that is common to a lot of the scaling papers \cite{GurYianilos, LSW, gurvits2004classical, garg2016deterministic, burgisser2017alternating, franks2018operator, burgisser2018efficient}. We will need the following lemma for \cite{LSW} which is essentially a robust version of the AM-GM inequality.

\begin{lem}\label{lem:LSW} Let $x_1,\ldots, x_n$ be positive real numbers s.t. $\sum_{i=1}^n x_i = n$ and $\sum_{i=1}^n (x_i - 1)^2 = \delta \le 1$. Then,
$$
\prod_{i=1}^n x_i \le \exp(-\delta/6)
$$
\end{lem}

\cref{lem:LSW} implies that for $\widehat{h}$ as in \cref{alg:template-scaling},
\begin{align}
\Det\left( \widehat{h}\right)^{1/n'} \ge \exp(-\eps'/12n') \label{eqn:ankit11}
\end{align}

Now the three steps of the analysis are as follows. The potential function is $\Phi(A) = |P(A)|^{1/\ell}$, and
its explicit description will be given in the next subsections.

\begin{enumerate}
\item {\bfseries Lower bound}: Since $P(A) \neq 0$, $P$ is a homogeneous polynomial of degree $\ell$ with integer coefficients and $A$ has rational entries with bit complexity at most $b$, it follows that $|P(A)| \ge 2^{-b\ell}$ and hence $\Phi(A) \ge 2^{-b}$.
\item {\bfseries Progress per step}: As long as $\widetilde{\ds}\left(A^{(t)} \right) \ge \eps$, 
$$
\Phi\left(A^{(t+1)} \right) \ge \exp(\eps'/12) \: \Phi\left(A^{(t)} \right)
$$
This follows from the invariance property of $P$ (under the action of $G$) and \cref{eqn:ankit11}. Since $P$ is invariant under the action of $G$, it follows that 
$$
P(h \cdot A) = \Det\left(\widehat{h}\right)^{\ell/n'} \: P(A)
$$ 
for all $h \in H$ (here $\widehat{h}$ is the only non-identity matrix in the tuple $h$).
\item {\bfseries Upper bound}: $\Phi\left(A^{(t)}\right) \le U$ for all $t$ because of \cref{eqn:ankit10} and due to the fact that $\norm{A^{(t)}} = 1$ for all $t$.
\end{enumerate}

The above three steps imply \cref{thm:unified}. It is quite magical that these invariant polynomials end up being useful potential functions for the analysis of these scaling algorithms. Without realizing the group actions at play, it might have been quite challenging to come up with potential functions for operator and tensor scaling algorithms (matrix scaling was done in \cite{LSW} without realizing the invariant theoretic connection). We also remark that most of the previous works use certain optimization problems called capacity (related to \cref{eqn:capacity}) as potential functions but invariant polynomials lie at the heart of the analysis.

The only thing left to complete the analysis is to get a handle on $U$ in \cref{eqn:ankit10}. This is something that needs to be done differently for different scaling problems and we proceed to do this next.

\subsection{Potential functions for matrix scaling}

Here the polynomials $P$ are extremely simple. We know that $A$ is not in the null cone iff the bipartite graph defined by $\text{supp}(A)$ has a perfect matching (\cref{prp:HM_matrix_scaling} and the succeeding discussion). Hence one can just take $P(A) = \prod_{i=1}^n A_{i,\sigma(i)}$ for an appropriate permutation $\sigma \in S_n$. This polynomial satisfies \cref{eqn:ankit10} with $U = 1$. We then get \cref{thm:matrix-scaling} from \cref{thm:unified}.\footnote{There is a slight discrepancy in parameters. This is because in \cref{alg:template-scaling}, we started already with normalized $A$'s.}

\subsection{Potential functions for operator scaling}

For the group action corresponding to operator scaling, i.e. left-right action (see \cref{subsubsec:op_scaling}), there is an explicit description of invariants.

\begin{thm}[\cite{DW2000, DZ2001, SVdB2001, ANS10}]\label{invariants-left-right}
The invariant ring of polynomials of the left-right action is generated by all polynomials of the form $\Det \left(\sum_{i=1}^m D_i \otimes A_i \right)$, where all $D_i$'s are $k \times k$ matrices and $k$ varies over $\NN$.
\end{thm}

This implies the following,

\begin{cor} $A = (A_1,\ldots, A_m)$ is scalable iff there exists $k \in \NN$ and $k \times k$ matrices $D_i$'s s.t.
$$
P(A) = \Det \left(\sum_{i=1}^m D_i \otimes A_i \right) \neq 0
$$
\end{cor}

Through an appropriate application of Alon's combinatorial nullstellansatz \cite{Alon_CN}, the AM-GM inequality and Cauchy-Schwarz inequality, the following can be proved \cite{garg2016deterministic},

\begin{cor} $A = (A_1,\ldots, A_m)$ is scalable iff there exists $k \in \NN$ and $k \times k$ integer matrices $D_i$'s s.t.
$$
P(A) = \Det \left(\sum_{i=1}^m D_i \otimes A_i \right) \neq 0
$$
and also,
$$
|P(A)| \le n^{n k/2} \norm{A}^{nk}
$$
\end{cor}

Since the degree is $nk$, we get that $U = \sqrt{n}$ in this case.\footnote{While a similar bound on $U$ follows from the discussion for tensor scaling (after all operator scaling is a special case), the method described here is more explicit.} Hence we get \cref{alg:operator-scaling} from \cref{thm:unified}.

\subsection{Potential functions for tensor scaling}

For tensor actions an explicit descriptions of the invariants is not known (for examples it is not known if there is a basis of invariant polynomials which are efficiently computable). However a semi-explicit description is known that can be used to bound $U$ by $n_1 n_2 \cdots n_d$. See \cite{burgisser2017alternating} for details. Hence one gets \cref{thm:alg-tensor} from \cref{thm:unified}.

We want to remark that the bounds on $U$ for operator and tensor scaling cases use sophisticated methods from invariant theory. Somewhat naive methods, e.g. reducing the problem to bounds on solutions to linear systems, only yield bounds on $|P(A)|$ which are doubly exponential in the degree $\ell$ (as opposed to singly exponential $\ell$) which would be useless in the tensor case since we can only bound the required degree by an exponential in the dimensions \cite{derksen2001polynomial}. For the left-right action though, $\ell$ can be assumed to be at most $n-1$ due to the work of Derksen and Makam \cite{derksen2015}. 

\section{Applications of scaling}\label{sec:apps}

As we saw in the previous chapters, scaling problems have a surprising connection to invariant theory,
which turns out to be fundamental to the analysis of the alternate minimization
algorithms which solve the scaling problems. This connection naturally leads to new and efficient algorithms
for problems in invariant theory. In this section we will see even more applications of scaling problems
in different areas of science.
For a complete discussion of the applications of matrix, operator and tensor scaling, we refer the 
reader to the papers~\cite{SZ90, I16, garg2016deterministic, garg2017algorithmic, burgisser2017alternating}.

\subsection{Matrix Scaling}\label{sec:ms-applications}

The matrix scaling problem has been posed (sometimes independently) in many different areas of study, 
ranging from telephone forecasting~\cite{K37}, economics~\cite{S62}, 
statistics~\cite{Sink}, image reconstruction algorithms~\cite{HL76}, linear algebra~\cite{FLS88, BDWY11}, 
optimization~\cite{ROTHBLUM1989737} and theoretical computer science~\cite{LSW, BDWY11}. 
For a more comprehensive list of references and historical overview of scaling problems, we recommend 
the survey~\cite{I16}, the papers~\cite{ROTHBLUM1989737, SZ90} and references therein. 
In this section we will describe three of the applications of matrix scaling cited above, providing a peek
on the abundance of applications of this simple and natural problem.

\paragraph{Computer Science:} Given a non-negative matrix $A \in \mat_n(\RR)$, its permanent is given by
the following expression: 
$$ \perm(A) = \sum_{\sigma \in S_n} \prod_{i=1}^n A_{i ,\sigma(i)}, $$
where $S_n$ is the group of permutations of the set $\{1,2 , \ldots, n\}$. This polynomial is extremely 
important in computer science, due to its completeness for a number of complexity classes. 

Computing the permanent of a 0-1 matrix is a $\csP$-complete problem, as shown by Valiant 
in~\cite{V79perm-comp}. 
Thus, much research has been devoted to computing a multiplicative approximation to the permanent, as 
described in~\cite{LSW} and references therein. 

Given a matrix $A \in \mat_n(\RR)$, note that the permanent of any scaling $BAC$ of $A$ is given by
$$ \perm(BAC) = \prod_{i=1}^n B_{i,i}C_{i,i} \cdot \perm(A). $$
Thus, if we could find diagonal matrices $B, C$ for which we knew a good approximation for $\perm(BAC)$,
the equality above would give us a good approximation for $\perm(A)$. As it turns out, for doubly-stochastic
matrices, good lower and upper bounds on the permanent are known! If $D$ is a doubly-stochastic matrix, 
the upper bound $\perm(D) \leq 1$ is trivial. For the lower bound, the solution to van der Waerden's conjecture
gives us that $\perm(D) \geq \dfrac{n!}{n^n} \geq e^{-n}$ \cite{friedland1979lower, falikman1981proof, egorychev1981solution, gurvits2004van, gurvits2008van}. Thus, finding a scaling of $A$ to a doubly stochastic
matrix gives us an $e^n$ approximation to compute the permanent! This approximation was given in the work
of Linial et al.~\cite{LSW}. While a fully polynomial randomized approximation scheme (FPRAS) is known for approximating the permanent \cite{jerrum2004polynomial}, the above algorithm, despite giving a much worse approximation (still a non-trivial one), is deterministic. The current best deterministic algorithm for approximating the permanent is due to Gurvits and Samorodnitsky \cite{gurvits2014bounds} (their paper also uses matrix scaling!), which achieves an approximation factor of $2^n$.


\paragraph{Combinatorial Geometry:} The Sylvester-Gallai theorem states that if $m$ distinct points
$p_1, \ldots, p_m \in \RR^n$ are arranged such that for any two distinct points $p_i, p_j$, there
exists a third point $p_k$ on the line defined by $p_i$ and $p_j$, then it must be the case that all 
points are collinear, that is, lie in a 1-dimensional subspace of $\RR^n$. 
This basic theorem in combinatorial geometry has many variants and generalizations,
which can be found in~\cite{BM90, BDWY11, DGOS} and references therein. 

A more quantitative version of
this problem, known as the $\delta$-SG problem, is defined as follows: if we now assume that the $m$ 
distinct points (could take them in $\CC^n$) are arranged such that for any point $p_i$, there are at least $\delta m$ points $p_j$
such that the line through $p_i$ and $p_j$ contains a third point $p_k$, can we say that these points
lie in a low dimensional subspace? Note that the original Sylvester-Gallai theorem is a special case
when $\delta = 1$. As it turns out, this $\delta$-SG problem can be phrased as a problem in linear algebra:
given a matrix $P \in \CC^{m \times n}$ whose rows are given by the points $p_1, \ldots, p_m$ satisfying
the arrangement contraints, is $P$ a low rank matrix? 

One approach to prove that such a matrix $P$ is low
rank is to find a high rank matrix $A \in \CC^{\ell \times m}$ such that $AP = 0$. In the case of the
$\delta$-SG problem, a natural matrix suggests itself: take $A$ to be the matrix which characterizes
the dependencies of the points $p_1, \ldots, p_m$. That is, for each triple $(i,j,k)$ such that 
$p_i, p_j, p_k$ are collinear, simply add a row to $A$ which encodes the linear combination of 
$p_i, p_j, p_k$ which gives zero. As it will soon be clear, this matrix $A$ which arises in the
$\delta$-SG problem is a very special type of matrix, and matrix scaling helps prove that such
matrices are always of high rank.

A surprising application of matrix scaling arises 
when one tries to obtain lower bounds on the rank of special types of matrices, called {\em design matrices},
which are defined based on the pattern of zero/non-zero entries in the matrix. More precisely, we say that 
a matrix $B \in \CC^{\ell \times m}$ is a $(q, k, t)$-design matrix if each row of $B$ has at most $q$ non-zero 
entries, each column has at least $k$ non-zero entries, and the supports of any two columns intersect in
at most $t$ rows. Note that the matrix $A$ from the $\delta$-SG problem is an example of a design matrix.
In~\cite{BDWY11}, the authors used matrix scaling to prove that any $\ell \times m$ matrix, 
where $\ell \geq m$, which is a $(q, k, t)$-design matrix has rank at least $m - \left(\dfrac{qtm}{2k}\right)^2$.
With this bound, they proved that any $\delta$-SG configuration must be in a subspace of dimention at most
$13/\delta^2$. This bound was improved in \cite{dvir2014improved} to $12/\delta$ by giving better bounds on ranks of design matrices (again relying heavily on matrix scaling).


\paragraph{Statistics:} It turns out that matrix scaling has an equivalent formulation as an entropy optimization problem, thereby
being very useful in statistics. 

\begin{prb}[Matrix Scaling - entropy formulation] Given two probability distributions $r, c \in \RR^n$ and a 
non-negative matrix $A \in \mat_n(\RR)$, find a non-negative matrix $B^*$ s.t. \footnote{$D(B||A) = \sum_{i,j} B_{i,j} \log(B_{i,j}/A_{i,j})$ is the KL-divergence between $B$ and $A$.}
\begin{align*} 
	B^* = \: &\textnormal{argmin} \: D(B || A) \: \text{s.t.} \\
	&\sum_{j=1}^n B_{ij} = r_i, \forall i \in [n] \\
	&\sum_{i=1}^n B_{ij} = c_j, \forall j \in [n] 
\end{align*}
\end{prb}


It turns out that the above optimization problem is equivalent to a non-uniform version of the matrix scaling problem, where one wants to scale $A$ to achieve marginals $r,c$ (as opposed to all $1$'s). The above optimization problem is trying to recover a joint probability distribution $B^*$ based on knowledge of the marginals $r,c$ (w.r.t. to some initial distribution $A$). Such estimation of joint probability distributions from partial data is abundant in statistics, as pointed
out by~\cite{SZ90}, with examples coming from estimating contingency tables, interregional migration, deriving
probability estimates from census data, and many others.

\subsection{Operator scaling}\label{sec:os-applications}

\paragraph{Non-Commutative Algebra and derandomizaton:} Whenever a certain mathematical object can be represented in several 
equivalent ways, a natural question which arises is the so called {\em word problem}: given two
representations of a mathematical object, do they describe the same object? Word problems are fundamental 
across many subareas of mathematics. In non-commutative algebra, when defining the {\em free skew field}, which is the 
field given by all rational functions over non-commutative variables (that is, the non-commutative
equivalent of the field of rational functions), the word problem arises as a computational problem
in a very natural way, which we describe next.
 
By the foundational work of Amitsur~\cite{A66}, the elements of the free skew field can be described by
equivalence classes of
arithmetic formulas, which take non-commutative variables and elements of the base field as inputs
and use linear combinations, multiplications and inverse gates to compute non-commutative rational expressions.
Two rational expressions are said to be equivalent if they have the same evaluation\footnote{whenever they
are defined in the given inputs, that is, the formula does not invert a singular matrix.} when we substitute 
the non-commuting variables by $d \times d$ matrices, for every $d \in \NN$. 

Therefore, we can phrase the word problem for the free skew field in a natural computational way: given
two non-commutative arithmetic formulas with inversion gates, are they equivalent (i.e. do they compute
the same rational function)? Note that Amitsur's work still leaves open the decidability of the word problem 
for the free skew field, as he provided no bounds on the dimension $d$. To prove that this word problem 
is decidable, in a series of works~\cite{C71, C73, C75}, Cohn reduced the word problem above to the problem of 
{\em non-commutative singularity
testing:} given a linear symbolic matrix $L = A_1 x_1 + \cdots + A_m x_m$, where $A_i \in \mat_n(\FF)$ and 
$x_i$'s are non-commutative variables, is the matrix $L$ singular over the free skew field? In this series of works,
Cohn essentially proved that any non-commutative formula can be computed by the (1,1) entry of the inverse of
such a symbolic matrix, which is the non-commutative analog of Valiant's completeness of determinant for commutative
formulas~\cite{V79comp-algebra}.

The connection between the non-commutative singularity problem and operator scaling comes from a theorem of
Cohn~\cite{Cohn95} which establishes that the symbolic matrix $L$ above is singular over the free skew field if,
and only if, the tuple of matrices $(A_1, \ldots, A_m)$ is dimension non-decreasing 
(see Definition~\ref{dfn:dim_non-increasing}). Therefore, by Theorem~\ref{thm:scalability_operator}, given
a symbolic matrix, to test its singularity we only need to test whether the tuple $(A_1, \ldots, A_m)$ can
be scaled to doubly-stochastic!

The above non-commutative singularity problem has a more well known cousin, namely the commutative singularity problem, more widely known as Edmonds' problem. Given a linear symbolic matrix $L = A_1 x_1 + \cdots + A_m x_m$, where $A_i \in \mat_n(\FF)$ and 
$x_i$'s are commutative variables, is the matrix $L$ singular over the field of rational functions? This problem has an easy randomized algorithm \cite{Lov1989}: plugging in random values and checking for singularity over the base field\footnote{In the problem, field is assumed to be large enough.} and it captures the famed polynomial identity testing problem through Valiant's completeness of the determinant \cite{V79comp-algebra}. A deterministic polynomial time algorithm for Edmonds' problem is a major open problem (with consequences for lower bounds in complexity theory \cite{KabImp}) and the scaling framework solves a closely related non-commutative cousin (over complex numbers or its subfields). In fact, Gurvits' original motivation to introduce operator scaling was to solve special cases of the Edmonds' problem. After \cite{garg2016deterministic}, the papers \cite{ivanyos2017noncommutative, derksen2015, ivanyos2017constructive} designed completely different and algebraic deterministic polynomial time algorithms for the non-commutative singularity problem which work over finite fields as well.\footnote{These papers for non-commutative singularity generalize the work of Raz and Shpilka \cite{RazShp} on polynomial identity testing of non-commutative algebraic programs which is a special case.} The work of \cite{BlaserJP16} designed a deterministic PTAS for the search version of the Edmonds' problem, partially inspired by the use Wong sequences in the algorithms of \cite{ivanyos2017noncommutative, ivanyos2017constructive}. It is an intriguing possibility that invariant theory and the methods surrounding it will have more to say about Edmonds' problem and in general, derandomization.

\paragraph{Invariant Theory:} As we saw in Section~\ref{sec:inv-theory}, a fundamental problem in 
computational invariant theory is the {\em null-cone problem}: given a group $G$ acting on a vector space 
$V$, and a point $v \in V$, decide whether $v$ is in the null cone $\cN_G(V)$. 

As it was also discussed in Section~\ref{sec:inv-theory}, the operator
scaling problem corresponds to the null-cone problem for the left-right action. Thus, the scaling algorithm
and its analysis prove that for this particular action the null-cone problem is in $\cP$. This was the first
polynomial time algorithm for the null-cone problem for the left-right action.

\paragraph{Combinatorial Geometry:} As we have seen in Section~\ref{sec:ms-applications}, matrix scaling
is very useful to prove rank bounds for design matrices, and such bounds found applications in 
combinatorial geometry. In very recent work~\cite{DGOS}, the authors generalize the definition of a 
design matrix to block matrices, which we will soon define, and used operator scaling to prove rank
bounds for design block matrices. 

These bounds were then used to obtain three new applications in combinatorial
geometry: bounding the projective rigidity of a configuration of points, obtaining
tight bounds on a generalization of the quantitative Sylvester-Gallai problem seen in 
Section~\ref{sec:ms-applications} and upper bounding the dimension of spaces containing certain 
configurations of low degree curves with many incidences, which can be seen as a variant of the
Sylvester-Gallai theorem to low degree curves. 

We say that a matrix $A$ is an $\ell \times m$ block matrix, with $d \times d$ blocks, if $A$ is a matrix 
whose entries $A_{ij}$ are matrices of dimension $d \times d$. When $d = 1$, we obtain the usual definition
of an $\ell \times m$ matrix. For a block matrix $A$, we denote its rank to be the rank of the $\ell d \times md$ 
matrix $\tilde{A}$ obtained from $A$ by ignoring the block structure. With this definition in mind, we 
can define {\em design block matrices} simply as follows: $A$ is a $(q, k, t)$-design block matrix if 
each row of $A$ has at most $q$ non-zero blocks, each column of $A$ contains at least $k$ non-singular blocks
and for any two columns, their support intersects in at most $t$ rows. In~\cite{DGOS}, the authors obtain
rank bounds for block design matrices, generalizing the results obtained for design matrices and 
discussed in Section~\ref{sec:ms-applications}. Operator scaling played a crucial rule in the proof. Note that a design block matrix cannot be thought of as a design matrix with some small blowup in parameters, since the columns corresponding to the same block could intersect at a lot of places, and hence the bounds of \cite{BDWY11, dvir2014improved} are not applicable in this setting. 

%
%

\subsection{Brascamp-Lieb inequalities and polytopes}\label{sec:bl-applications}

Another important application of operator scaling appears in functional analysis and optimization, 
towards the celebrated {\em Brascamp-Lieb inequalities}~\cite{BL76, Lieb90} and their corresponding 
{\em Brascamp-Lieb polytopes}. The Brascamp-Lieb inequalities (BL for short) and their reverse form
generalize many important inequalities, such as Cauchy-Schwarz and H\"{o}lder's inequalities, 
Loomis-Whitney inequality, Young's convolution inequalities and many others. In this section, we will
describe how BL-inequalities can be seen as a particular case of the operator scaling problem, and
discuss some applications in combinatorics and complexity. For a more in depth discussion of BL inequalities,
we refer the reader to the papers~\cite{BCCT, garg2017algorithmic} and references therein.

A BL datum is given by a tuple of matrices $\bB = (B_1, \ldots, B_m)$ where $B_i \in \RR^{n_i \times n}$
and a tuple of non-negative reals $\bp = (p_1, \ldots, p_m)$. We will represent a BL datum by the tuple 
$(\bB, \bp)$. The BL inequality with datum $(\bB, \bp)$ states that for every tuple of non-negative, Lebesgue
integrable functions $(f_1, \ldots, f_m)$, where $f_i : \RR^{n_i} \rightarrow \RR$, the following inequality
holds:\footnote{Below, $\| f_i \|_{1/p_i} = \left(\int_{y_i \in \RR^{n_i}} f_i(y_i)^{1/p_i} dy_i\right)^{p_i}$.}
\begin{equation}
	\int_{x \in \RR^n} \prod_{i=1}^m f_i(B_i x) dx \leq C \cdot \prod_{i=1}^m \| f_i \|_{1/p_i},
\end{equation}
for some constant $C \in (0, \infty]$ which is independent of the functions $f_i$. When $C$ is finite (which
is when we indeed have a non-trivial inequality), we say that the datum $(\bB, \bp)$ is feasible,
and denote by $BL(\bB, \bp)$ the smallest value of $C$ for which the inequality always holds, 
which we refer to as the BL constant.

Ball~\cite{B89} and Barthe~\cite{B98} proved that for certain types of BL data, the BL constant will always
equal 1. They called these types of BL data {\em geometric}, which we now define:

\begin{dfn}[Geometric BL datum]\label{dfn:geometric} 
A BL datum is called {\em geometric} if it satisfies the following conditions:
\begin{enumerate}
	\item \textbf{Isotropy:} $\sum_{i=1}^m p_i B_i^TB_i = I_n$.
	\item \textbf{Projection:} For every $i \in [m]$, $B_i$ is a projection matrix, that is, 
	$B_iB_i^T = I_{n_i}$.
\end{enumerate}	
\end{dfn}  

Note that the definition above is remarkably similar to the definition of doubly stochastic operators
(Definition~\ref{dfn:doubly_stochastic_op}), and this similarity is an important part of the connection
between BL inequalities and operator scaling.

An important characterization of feasible BL data was given by Bennet et al.~\cite{BCCT}, and this 
characterization, as we will soon see, is remarkably similar to the dimension non-decreasing definition
(Definition~\ref{dfn:dim_non-increasing}).

\begin{thm}[Feasibility of BL datum~\cite{BCCT}]\label{thm:feasibility} 
The datum $(\bB, \bp)$ is feasible iff the following inequalities hold:
\begin{enumerate}
	\item $n = \sum_{i=1}^m p_i n_i$
	\item $\dim(V) \leq \sum_{i=1}^m p_i \dim(B_i (V)),$ for all subspaces $V \subseteq \RR^n$.
\end{enumerate} 
\end{thm}

Note that for a given tuple of matrices $\bB$, the theorem above proves that the set of vectors $\bp$
for which the datum $(\bB, \bp)$ is feasible is given by a polytope.\footnote{One can see that the number
of inequalities is finite because given a tuple $\bB$, the numbers $\dim(B_i V)$ lie in the set $[n]$, 
therefore giving us at most $n^m$ different inequalities.} This polytope, which we denote by $P_\bB$, is
the so called BL polytope.

In~\cite{garg2017algorithmic}, the authors reduce the feasibility of BL data to an operator scaling problem,
thereby giving an efficient algorithm for the membership problem in a BL polytope (that is, solving the
problem: given datum $(\bB, \bp)$, is $\bp \in P_\bB$?). We refer the reader to \cite{garg2017algorithmic} for details (or perhaps find the reduction yourself!).

The reduction from BL to operator scaling along with \cref{thm:scalability_operator} yields the following BL scaling theorem (which is already present in the work of \cite{BCCT}).

\begin{thm}\label{thm:bl_scalability}
A BL datum $(\bB, \bp)$ is feasible iff there are invertible matrices (denoted {\em BL scalings}) 
$A \in \GL_n(\RR)$, $C_i \in \GL_{n_i}(\RR)$
such that the datum $(\bB', \bp)$ is geometric, where $\bB' = (C_1B_1A, \ldots, C_mB_mA)$.\footnote{The
precise statement would be that the new datum $(\bB', \bp)$ is "{\em close to} geometric," which is
analogous to what happens in the operator scaling setting. For simplicity, we forgo the exact statement.}
\end{thm}

With the definitions and results above, we are ready to discuss some applications of BL inequalities and polytopes.

\paragraph{Complexity Theory:} Forster's celebrated lower bound on the unbounded error 
probabilistic communication complexity or the sign rank uses a 
remarkable result, proved by him in~\cite[Theorem 4.1]{F02}, 
which can be stated as follows: 

\begin{thm}[\cite{F02}] Let $v_1, \ldots, v_m \in \RR^n$, where $m \geq n$, be a set of vectors in general
position, that is, any subset of $n$ of these vectors are linearly independent. Then, there exists a 
matrix $A \in \GL_n(\RR)$ such that the following holds:
\begin{equation}\label{eq:forster}
	\sum_{i=1}^m \dfrac{n}{m} \cdot \dfrac{(Av_i)(Av_i)^T}{\| Av_i \|_2^2} = I_n 
\end{equation} 
\end{thm} 

The condition given above, when stated in the language of BL inequalities, becomes exactly
the claim that the BL datum $(\bB, \bp)$ given by $B_i = v_i^T$ and $p_i = n/m$ for $i \in [m]$, can be 
scaled to a geometric datum when the vectors $v_i$ are in general position! We note that generalizations of Forster's theorem already appear in two previous works, explicitly in \cite{gurvits2002deterministic} and implicitly in \cite{B98}.

Since Forster's result is a special case of Theorem~\ref{thm:bl_scalability}, which itself
is a special case of the operator scaling problem, we see that the operator scaling theory gives a vast 
generalization to Forster's theorem.


\paragraph{Combinatorial Optimization:} BL polytopes are interesting combinatorially because they can be very complex,
having exponentially many facets, while admitting a very succinct description (given by the tuple of matrices
$\bB$). The work in~\cite{garg2017algorithmic} gives a membership oracle, as well as a separation oracle
for BL polytopes\footnote{The running time of the oracles depends on the common denominator of the vector
$\bp$. For more details, we refer the reader to the paper.} and therefore these polytopes could be a useful 
tool for solving natural optimization problems. Thus, looking for natural polytopes which are special cases
of BL polytopes is a first step in understanding their expressive power, which is far from understood.

One polytope which can be encoded as a BL polytope in a simple way is the linear matroid intersection polytope,
which we now describe. The linear matroid associated with a tuple of vectors 
$v = (v_1, \ldots, v_m)$, where $v_i \in \RR^n$, 
is the matroid with the following collection of independent sets: 
$\cM_v = \{ I \subseteq [m] \mid (v_i)_{i \in I} \text{ are linearly independent}  \}$.

Given two tuples of vectors $v = (v_1, \ldots, v_m)$ and $w = (w_1, \ldots, w_m)$, defining two linear
matroids $\cM_v, \cM_w$ over $\RR^n$, their (linear matroid) intersection polytope is given by the convex 
hull of the characteristic vectors of their common independent sets. That is,
$$ P_{\cM_v, \cM_w} = \conv\{ 1_I \mid I \subseteq [m] \text{ s.t. } (v_i)_{i \in I} 
\text{ and } (w_i)_{i \in I} \text{ are linearly independent} \}. $$

In~\cite{garg2017algorithmic}, the authors prove\footnote{There was a mistake in their original proof, which
was fixed thanks to Damien Strazak and Nisheeth Vishnoi.} that the polytope $P_{\cM_v, \cM_w}$ corresponds
to the BL polytope given by the matrices 
$B_i = \begin{pmatrix} 0 & v_i^T \\ w_i^T & 0 \end{pmatrix}$, where each $0$ corresponds to the zero vector
in $\RR^n$.

\subsection{Tensor scaling}\label{sec:ts-applications}

\paragraph{Entanglement distillation:} The tensor scaling problem has a very natural interpretation in
quantum information theory. If we regard the vector space $V = \Ten(n_1, \ldots, n_d)$ as the set of pure
states of a quantum system with $d$ particles,\footnote{We would have to consider only tensors of unit norm.} 
the scaling action of $G = \SL(n_1) \ot \cdots \ot \SL(n_d )$ corresponds to a class of quantum operations
called {\em stochastic local operations and classical communication (SLOCC)}, defined in~\cite{bennett2000exact}. These
operations on a quantum system have a natural communication complexity interpretation: each party is holding a particle of the system, parties are allowed free classical communication (i.e. sending
bits to one another) and each party can perform quantum operations and
measurements on its own particle, and finally we allow {\em post selection} on measurement outcomes. 



Quantum states with uniform marginals are called {\em locally maximally entangled} and hence the tensor scaling question is about distilling locally maximally entangled states from a given pure state by SLOCC operations.

\paragraph{Slice-Rank:} The slice-rank of a tensor, introduced in~\cite{T16}, is a different notion of 
tensor rank which has found applications in extremal combinatorics and number 
theory (for more details, see~\cite{BCC+17} and references therein). A tensor $B \in \Ten(n_1, \ldots, n_d)$
is said to have {\em slice-rank one} if it is the tensor product of a vector and a lower order tensor, that 
is, if there exists an index $j \in [d]$ such that $B = v \ot_j C$, where $v \in \CC^{n_j}$ and 
$C \in \Ten(n_1, \ldots, n_{j-1}, n_{j+1}, \ldots, n_d)$. With this definition, the {\em slice-rank}
of a tensor, $A \in \Ten(n_1, \ldots, n_d)$, is defined as the smallest $k$ for which $A$ can be decomposed as the 
sum of $k$ slice-rank one tensors. We denote the slice rank of $A$ by $\srk(A)$. 

Let us focus on the case $n_1 = \cdots = n_d = n$ (not all of what we are going to say holds in the unbalanced case). Note that the slice-rank of a tensor can be at most $n$, since we can always flatten
the tensor on any one coordinate, and the matrix rank decomposition of the flattening provides a slice decomposition of the tensor.
In a recent work~\cite{BCC+17}, the authors developed a connection between the slice rank of a tensor 
(and an asymptotic version of slice-rank) and the null cone of the tensor scaling group action. 
More precisely the authors proved the following two theorems:

\begin{thm}\label{thm:sr-nullcone}
	Given $A \in \Ten(n, \ldots, n)$, if $\srk(A) < n$, then $A$ is in the
	null cone of the tensor scaling action.
\end{thm}

\begin{thm}\label{thm:asymp-non-full}
	If a tensor $A \in \Ten(n, \ldots, n)$ is in the null cone of the tensor action, then there exists
	$k \in \NN$ such that $\srk(A^{\ot k}) < n^k$.
\end{thm}

Therefore deciding whether a particular tensor lies in the null cone of the tensor action could give us
information on its slice-rank or on the asymptotic version of slice-rank. Relying on \cref{thm:sr-nullcone,,thm:asymp-non-full}, 
it was proved in \cite{burgisser2017alternating} that being in the null cone of the tensor action and non-fullness of the asymptotic 
slice-rank are equivalent conditions (in other words Theorem~\ref{thm:asymp-non-full} is in fact
an equivalence). 


\section{Conclusion and open problems}\label{sec:conclusion}

We have seen that scaling problems are particular instances of fundamental problems in invariant theory, and that
invariant theory provides a rich source of potential functions to analyze the natural alternating
minimization algorithms for the scaling problems. Therefore, settling the complexity of problems in invariant
theory will have many applications not only in computational invariant theory or geometric complexity
theory, but also in many other areas of science. We believe that the recent series of works on scaling algorithms 
are only the beginning of many future discoveries, and to witness this we present several problems which are still
open in the area:

\begin{enumerate}
	\item Design a polynomial time algorithm for tensor scaling with a $\poly(\log(1/\eps))$ dependence on the error 	
	parameter $\eps$. This will yield a polynomial time algorithm for the null-cone problem for tensor actions. Such 				algorithms already exist for the matrix scaling \cite{KALANTARI199687, LSW, CohenMTV17, AllenZhuLOW17} and the 			operator scaling \cite{AGLOW18} problems. This involves exploring algorithms for geodesically convex optimization (see 			\cite{AGLOW18} and references therein for a discussion).

	\item Is there a polynomial time algorithm for the null-cone problem for more general group actions? The moment map 			and the optimization problem in \cref{eqn:capacity} provide an analytic approach to this. 

	\item Can one understand the behavior of \cref{alg:matrix-scaling,,alg:operator-scaling,,alg:tensor-scaling} when 				the object $A$ is not scalable? Does the algorithm converge to some cycle? How close does one get to satisfying the 				stochasticity constraints?
	
	\item Can we find more applications of scaling problems in computer science? More generally, can we find
	instances of the non-commutative duality appearing in computer science?

	\item As mentioned in Section~\ref{sec:analysis}, the analysis of the scaling algorithms rely on the existence
	of a generating set of invariant polynomials which is ``nice'' in the following sense: each polynomial in this
	generating set has ``small'' coefficients (of size exponential in the degree of the polynomials). Is there
	a systematic way of obtaining such nice set of polynomials for more general group actions?
\end{enumerate}

\section{Other recent scaling works}\label{sec:related-works}

This survey covered a particular family of scaling problems, which is usually referred to as {\em uniform scaling
problems}, as the matrix, operator and tensor scaling problems here defined only concern with the possibility
of scaling the input to a doubly-stochastic (matrix and operator scaling) or a $d$-stochastic element 
(tensor scaling). However, this is not the whole story, as one could ask the following question: given prescribed
marginals and an input matrix/operator/tensor, can we scale the input to have the prescribed marginals?

This more general question has also been studied extensively. In the matrix scaling case, the theory of non-uniform scaling is not much different from the theory of uniform scaling. However in the operator and tensor scaling settings, there are a lot more twists in the non-uniform case. Recent works have made remarkable progress towards this 
end~\cite{franks2018operator, burgisser2018efficient}. In this more general setting, more 
sophisticated concepts from invariant theory and representation theory are needed for the analysis of the algorithms.

Another line of research has been in the development of faster algorithms for scaling problems, with a different
approach than the alternating minimization described in this survey. Recently some remarkable successes have been
obtained in this direction, with nearly-linear algorithms being developed for 
matrix scaling~\cite{CohenMTV17, AllenZhuLOW17} and a faster algorithm ($\poly(\log(1/\eps))$ convergence rate as opposed to $\poly(1/\epsilon)$) being developed for the operator scaling
problem~\citep{AGLOW18}. 

Other variants of the scaling problems above have also been studied, and we cite here a few. 
The {\em matrix balancing} problem is a variant of matrix scaling where given a square matrix $A$ with 
complex entries, the goal is to decide whether there is a diagonal matrix $D$ such that $DAD^{-1}$ is 
{\em balanced}, that is, where the $i^{th}$ row has the same norm as the $i^{th}$ column. This problem 
also has applications in different areas of computer science and numerical analysis. For more details on 
matrix balancing, we refer the reader to the survey~\cite{I16} and references therein. 
In~\cite{KLLR17}, the authors solve
the Paulsen problem in operator theory and they use several tools from the theory of operator scaling, in addition to other sophisticated methods.

\section*{Other resources/pointers}

The following workshop page is a very useful resource (containing lecture videos and pointers to papers) for getting further into the topics related to this survey: \href{https://www.math.ias.edu/ocit2018}{Optimization, Complexity and Invariant Theory}.



\section*{Acknowledgements}
We thank Vikraman Arvind for inviting us to write this survey for the EATCS complexity column (June 2018 issue) and Avi Wigderson for providing helpful comments on an earlier version of this survey. We would also like to thank our collaborators Peter B\"{u}rgisser, Cole Franks, Michael Walter and Avi Wigderson for our many interesting discussions about the above problems.

\bibliographystyle{alphaurl}
\addcontentsline{toc}{section}{References}
\bibliography{survey}

\newcommand{\etalchar}[1]{$^{#1}$}
\begin{thebibliography}{AZGL{\etalchar{+}}18}

\bibitem[AFG14]{FG14}
Michael A.~Forbes and Venkatesan Guruswami.
\newblock Dimension expanders via rank condensers.
\newblock {\em arXiv preprint arXiv:1411.7455}, 2014.

\bibitem[Alo99]{Alon_CN}
Noga Alon.
\newblock Combinatorial {Nullstellensatz}.
\newblock {\em Combinatorics, Probability and Computing}, 8(1-2):7--29, 1999.

\bibitem[ALOW17]{AllenZhuLOW17}
Zeyuan Allen{-}Zhu, Yuanzhi Li, Rafael Oliveira, and Avi Wigderson.
\newblock Much faster algorithms for matrix scaling.
\newblock In {\em 58th {IEEE} Annual Symposium on Foundations of Computer
  Science, {FOCS} 2017, Berkeley, CA, USA, October 15-17, 2017}, pages
  890--901, 2017.
\newblock URL: \url{https://doi.org/10.1109/FOCS.2017.87}, \href
  {http://dx.doi.org/10.1109/FOCS.2017.87} {\path{doi:10.1109/FOCS.2017.87}}.

\bibitem[Ami66]{A66}
Shimshon Amitsur.
\newblock Rational identities and applications to algebra and geometry.
\newblock {\em Journal of Algebra}, 3:304--359, 1966.

\bibitem[ANS10]{ANS10}
Bharat Adsul, Suresh Nayak, and K.~V. Subrahmanyam.
\newblock A geometric approach to the kronecker problem ii : rectangular
  shapes, invariants of matrices, and a generalization of the {Artin}-{Procesi}
  theorem.
\newblock {\em Manuscript, available at http://www.cmi.ac.in/~kv/ANS10.pdf},
  2010.

\bibitem[AZGL{\etalchar{+}}18]{AGLOW18}
Zeyuan Allen-Zhu, Ankit Garg, Yuanzhi Li, Rafael Oliveira, and Avi Wigderson.
\newblock Operator scaling via geodesically convex optimization, invariant
  theory and polynomial identity testing.
\newblock {\em STOC}, 2018.

\bibitem[Bal89]{B89}
Keith Ball.
\newblock Volumes of sections of cubes and related problems.
\newblock {\em Geometric Aspects of Functional Analysis}, pages 251--260, 1989.

\bibitem[Bar98]{B98}
Franck Barthe.
\newblock On a reverse form of the {Brascamp}--{Lieb} inequality.
\newblock {\em Inventiones Mathematicae}, 134:335--361, 1998.

\bibitem[BCC{\etalchar{+}}17]{BCC+17}
Jonah Blasiak, Thomas Church, Henry Cohn, Joshua~A. Grochow, Eric Naslund,
  William~F Sawin, and Chris Umans.
\newblock On cap sets and the group-theoretic approach to matrix
  multiplication.
\newblock {\em Discrete Analysis}, 2017.
\newblock URL: \url{http://arxiv.org/abs/1605.06702}, \href
  {http://arxiv.org/abs/1605.06702} {\path{arXiv:1605.06702}}, \href
  {http://dx.doi.org/10.19086/da.1245} {\path{doi:10.19086/da.1245}}.

\bibitem[BCCT08]{BCCT}
Jonathan Bennett, Anthony Carbery, Michael Christ, and Terence Tao.
\newblock The {Brascamp}-{Lieb} inequalities: finiteness, structure, and
  extremals.
\newblock {\em Geometric and Functional Analysis}, 17(5):1343--1415, 2008.

\bibitem[BDYW11]{BDWY11}
Boaz Barak, Zeev Dvir, Amir Yehudayoff, and Avi Wigderson.
\newblock Rank bounds for design matrices with applications to combinatorial
  geometry and locally correctable codes.
\newblock In {\em Proceedings of the forty-third annual ACM symposium on Theory
  of computing}, pages 519--528. ACM, 2011.

\bibitem[BFG{\etalchar{+}}18]{burgisser2018efficient}
Peter B{\"u}rgisser, Cole Franks, Ankit Garg, Rafael Oliveira, Michael Walter,
  and Avi Wigderson.
\newblock Efficient algorithms for tensor scaling, quantum marginals and moment
  polytopes.
\newblock {\em arXiv preprint arXiv:1804.04739}, 2018.

\bibitem[BGO{\etalchar{+}}18]{burgisser2017alternating}
Peter B{\"u}rgisser, Ankit Garg, Rafael Oliveira, Michael Walter, and Avi
  Wigderson.
\newblock Alternating minimization, scaling algorithms, and the null-cone
  problem from invariant theory.
\newblock {\em Proceedings of Innovations in Theoretical Computer Science
  (ITCS)}, 2018.
\newblock \href {http://arxiv.org/abs/1711.08039} {\path{arXiv:1711.08039}}.

\bibitem[BI11]{burgisser2011geometric}
Peter B{\"u}rgisser and Christian Ikenmeyer.
\newblock Geometric complexity theory and tensor rank.
\newblock In {\em Proceedings of the Symposium on the Theory of Computing (STOC
  2011)}, pages 509--518. ACM, 2011.
\newblock \href {http://arxiv.org/abs/1011.1350} {\path{arXiv:1011.1350}},
  \href {http://dx.doi.org/10.1145/1993636.1993704}
  {\path{doi:10.1145/1993636.1993704}}.

\bibitem[BI13]{burgisser2013explicit}
Peter B{\"u}rgisser and Christian Ikenmeyer.
\newblock Explicit lower bounds via geometric complexity theory.
\newblock In {\em Proceedings of the Symposium on the Theory of Computing (STOC
  2013)}, pages 141--150. ACM, 2013.
\newblock \href {http://arxiv.org/abs/1210.8368} {\path{arXiv:1210.8368}},
  \href {http://dx.doi.org/10.1145/2488608.2488627}
  {\path{doi:10.1145/2488608.2488627}}.

\bibitem[BJP16]{BlaserJP16}
Markus Bl{\"{a}}ser, Gorav Jindal, and Anurag Pandey.
\newblock Greedy strikes again: {A} deterministic {PTAS} for commutative rank
  of matrix spaces.
\newblock {\em Electronic Colloquium on Computational Complexity {(ECCC)}},
  23:145, 2016.
\newblock URL: \url{http://eccc.hpi-web.de/report/2016/145}.

\bibitem[BL76]{BL76}
Herm Brascamp and Elliot Lieb.
\newblock Best constants in {Young's} inequality, its converse and its
  generalization to more than three functions.
\newblock {\em Advances in Mathematics}, 20:151--172, 1976.

\bibitem[BM90]{BM90}
Peter Borwein and William~OJ Moser.
\newblock A survey of sylvester's problem and its generalizations.
\newblock {\em Aequationes Mathematicae}, 40(1):111--135, 1990.

\bibitem[BPR{\etalchar{+}}00]{bennett2000exact}
Charles~H Bennett, Sandu Popescu, Daniel Rohrlich, John~A Smolin, and Ashish~V
  Thapliyal.
\newblock Exact and asymptotic measures of multipartite pure-state
  entanglement.
\newblock {\em Physical Review A}, 63(1):012307, 2000.

\bibitem[B{\"u}r12]{Bugisser_GCT}
Peter B{\"u}rgisser.
\newblock Prospects for geometric complexity theory.
\newblock In {\em 2012 IEEE 27th Conference on Computational Complexity}, pages
  235--235, June 2012.
\newblock \href {http://dx.doi.org/10.1109/CCC.2012.19}
  {\path{doi:10.1109/CCC.2012.19}}.

\bibitem[CMTV17]{CohenMTV17}
Michael~B. Cohen, Aleksander Madry, Dimitris Tsipras, and Adrian Vladu.
\newblock Matrix scaling and balancing via box constrained newton's method and
  interior point methods.
\newblock In {\em 58th {IEEE} Annual Symposium on Foundations of Computer
  Science, {FOCS} 2017, Berkeley, CA, USA, October 15-17, 2017}, pages
  902--913, 2017.
\newblock URL: \url{https://doi.org/10.1109/FOCS.2017.88}, \href
  {http://dx.doi.org/10.1109/FOCS.2017.88} {\path{doi:10.1109/FOCS.2017.88}}.

\bibitem[Coh71]{C71}
P.~M. Cohn.
\newblock The embedding of firs in skew fields.
\newblock {\em Proceedings of the London Mathematical Society}, 23:193--213,
  1971.

\bibitem[Coh73]{C73}
P.~M. Cohn.
\newblock The word problem for free fields.
\newblock {\em The Journal of Symbolic Logic}, 38(2):309--314, 1973.

\bibitem[Coh75]{C75}
P.~M. Cohn.
\newblock The word problem for free fields: A correction and an addendum.
\newblock {\em Journal of Symbolic Logic}, 40(1):69--74, 1975.

\bibitem[Coh95]{Cohn95}
Paul Cohn.
\newblock {\em Skew Fields, Theory of General Division Rings}, volume~57.
\newblock Encyclopedia of Mathematics and its Applications, 1995.

\bibitem[Der01]{derksen2001polynomial}
Harm Derksen.
\newblock Polynomial bounds for rings of invariants.
\newblock {\em Proceedings of the American Mathematical Society},
  129(4):955--963, 2001.
\newblock \href {http://dx.doi.org/10.1090/S0002-9939-00-05698-7}
  {\path{doi:10.1090/S0002-9939-00-05698-7}}.

\bibitem[DGOS16]{DGOS}
Zeev Dvir, Ankit Garg, Rafael Oliveira, and J{\'o}zsef Solymosi.
\newblock Rank bounds for design matrices with block entries and geometric
  applications.
\newblock {\em arXiv preprint arXiv:1610.08923}, 2016.

\bibitem[DK15]{derksen2015computational}
Harm Derksen and Gregor Kemper.
\newblock {\em Computational invariant theory}.
\newblock Springer, 2015.

\bibitem[DM15]{derksen2015}
Harm Derksen and Visu Makam.
\newblock Polynomial degree bounds for matrix semi-invariants.
\newblock 2015.
\newblock \href {http://arxiv.org/abs/1512.03393} {\path{arXiv:1512.03393}}.

\bibitem[DSW14]{dvir2014improved}
Zeev Dvir, Shubhangi Saraf, and Avi Wigderson.
\newblock Improved rank bounds for design matrices and a new proof of {Kelly}'s
  theorem.
\newblock In {\em Forum of Mathematics, Sigma}, volume~2. Cambridge University
  Press, 2014.

\bibitem[DW00]{DW2000}
Harm Derksen and Jerzy Weyman.
\newblock Semi-invariants of quivers and saturation for littlewood-richardson
  coefficients.
\newblock {\em Journal of the American Mathematical Society}, 13(3):467--479,
  2000.

\bibitem[DZ01]{DZ2001}
M{\'a}ty{\'a}s Domokos and A.~N. Zubkov.
\newblock Semi-invariants of quivers as determinants.
\newblock {\em Transformation Groups}, 6(1):9--24, 2001.

\bibitem[Ego81]{egorychev1981solution}
Gregory~P Egorychev.
\newblock The solution of van der waerden's problem for permanents.
\newblock {\em Advances in Mathematics}, 42(3):299--305, 1981.

\bibitem[Fal81]{falikman1981proof}
Dmitry~I Falikman.
\newblock Proof of the van der waerden conjecture regarding the permanent of a
  doubly stochastic matrix.
\newblock {\em Mathematical notes of the Academy of Sciences of the USSR},
  29(6):475--479, 1981.

\bibitem[FL89]{FRANKLIN1989717}
Joel Franklin and Jens Lorenz.
\newblock On the scaling of multidimensional matrices.
\newblock {\em Linear Algebra and its Applications}, 114-115:717 -- 735, 1989.
\newblock Special Issue Dedicated to Alan J. Hoffman.
\newblock URL:
  \url{http://www.sciencedirect.com/science/article/pii/0024379589904904},
  \href {http://dx.doi.org/https://doi.org/10.1016/0024-3795(89)90490-4}
  {\path{doi:https://doi.org/10.1016/0024-3795(89)90490-4}}.

\bibitem[FLS88]{FLS88}
Shmuel Friedland, Chi-Kwong Li, and Hans Schneider.
\newblock Additive decomposition of nonnegative matrices with applications to
  permanents and scalingt.
\newblock {\em Linear and Multilinear Algebra}, 23(1):63--78, 1988.

\bibitem[For02]{F02}
J{\"u}rgen Forster.
\newblock A linear lower bound on the unbounded error probabilistic
  communication complexity.
\newblock {\em Journal of Computer and System Sciences}, 65(4):612--625, 2002.

\bibitem[Fra18]{franks2018operator}
Cole Franks.
\newblock Operator scaling with specified marginals.
\newblock {\em STOC}, 2018.
\newblock \href {http://arxiv.org/abs/1801.01412} {\path{arXiv:1801.01412}}.

\bibitem[Fri79]{friedland1979lower}
Shmuel Friedland.
\newblock A lower bound for the permanent of a doubly stochastic matrix.
\newblock {\em Annals of Mathematics}, pages 167--176, 1979.

\bibitem[GGOW16]{garg2016deterministic}
Ankit Garg, Leonid Gurvits, Rafael Oliveira, and Avi Wigderson.
\newblock A deterministic polynomial time algorithm for non-commutative
  rational identity testing.
\newblock In {\em Proceedings of the Symposium on Foundations of Computer
  Science (FOCS 2016)}, pages 109--117. IEEE, 2016.
\newblock \href {http://arxiv.org/abs/1511.03730} {\path{arXiv:1511.03730}},
  \href {http://dx.doi.org/10.1109/FOCS.2016.95}
  {\path{doi:10.1109/FOCS.2016.95}}.

\bibitem[GGOW17]{garg2017algorithmic}
Ankit Garg, Leonid Gurvits, Rafael Oliveira, and Avi Wigderson.
\newblock Algorithmic and optimization aspects of brascamp-lieb inequalities,
  via operator scaling.
\newblock In {\em Proceedings of the Symposium on the Theory of Computing (STOC
  2017)}, pages 397--409. ACM, 2017.
\newblock \href {http://arxiv.org/abs/1607.06711} {\path{arXiv:1607.06711}}.

\bibitem[GRS13]{georgoulas2013moment}
Valentina Georgoulas, Joel~W Robbin, and Dietmar~A Salamon.
\newblock The moment-weight inequality and the {H}ilbert-{M}umford criterion.
\newblock {\em \href{https://arxiv.org/abs/1311.0410}{arXiv:1311.0410}}, 2013.

\bibitem[GS02]{gurvits2002deterministic}
Leonid Gurvits and Alex Samorodnitsky.
\newblock A deterministic algorithm for approximating the mixed discriminant
  and mixed volume, and a combinatorial corollary.
\newblock {\em Discrete \& Computational Geometry}, 27(4):531--550, 2002.

\bibitem[GS14]{gurvits2014bounds}
Leonid Gurvits and Alex Samorodnitsky.
\newblock Bounds on the permanent and some applications.
\newblock In {\em Foundations of Computer Science (FOCS), 2014 IEEE 55th Annual
  Symposium on}, pages 90--99. IEEE, 2014.

\bibitem[Gur04a]{gurvits2004classical}
Leonid Gurvits.
\newblock Classical complexity and quantum entanglement.
\newblock {\em Journal of Computer and System Sciences}, 69(3):448--484, 2004.
\newblock \href {http://dx.doi.org/10.1016/j.jcss.2004.06.003}
  {\path{doi:10.1016/j.jcss.2004.06.003}}.

\bibitem[Gur04b]{gurvits2004van}
Leonid Gurvits.
\newblock Van der waerden conjecture for mixed discriminants.
\newblock {\em arXiv preprint math/0406420}, 2004.

\bibitem[Gur08]{gurvits2008van}
Leonid Gurvits.
\newblock Van der waerden/schrijver-valiant like conjectures and stable (aka
  hyperbolic) homogeneous polynomials: one theorem for all.
\newblock {\em the electronic journal of combinatorics}, 15(1):66, 2008.

\bibitem[GY98]{GurYianilos}
Leonid Gurvits and Peter~N. Yianilos.
\newblock The deflation-inflation method for certain semidefinite programming
  and maximum determinant completion problems.
\newblock {\em Technical Report, NECI}, 1998.

\bibitem[HH12]{Hechman-Hochs12}
Gert Heckman and Peter Hochs.
\newblock Geometry of the momentum map, 2012.
\newblock URL:
  \url{http://www.maths.adelaide.edu.au/peter.hochs/momentum_new.pdf}.

\bibitem[Hil90]{Hil90}
David Hilbert.
\newblock Ueber die {T}heorie der algebraischen {F}ormen.
\newblock {\em Mathematische Annalen}, 36(4):473--534, 1890.

\bibitem[Hil93]{Hil}
David Hilbert.
\newblock Uber die vollen {I}nvariantensysteme.
\newblock {\em Math. Ann.}, 42:313--370, 1893.

\bibitem[HL76]{HL76}
Gabor~T Herman and Arnold Lent.
\newblock Iterative reconstruction algorithms.
\newblock {\em Computers in biology and medicine}, 6(4):273--294, 1976.

\bibitem[Ide16]{I16}
Martin Idel.
\newblock A review of matrix scaling and sinkhorn's normal form for matrices
  and positive maps.
\newblock {\em arXiv preprint arXiv:1609.06349}, 2016.

\bibitem[IQS17a]{ivanyos2017constructive}
G{\'a}bor Ivanyos, Youming Qiao, and KV~Subrahmanyam.
\newblock Constructive non-commutative rank computation is in deterministic
  polynomial time.
\newblock {\em Proceedings of Innovations in Theoretical Computer Science
  (ITCS)}, 2017.
\newblock \href {http://arxiv.org/abs/1512.03531} {\path{arXiv:1512.03531}},
  \href {http://dx.doi.org/10.4230/LIPIcs.ITCS.2017.55}
  {\path{doi:10.4230/LIPIcs.ITCS.2017.55}}.

\bibitem[IQS17b]{ivanyos2017noncommutative}
G{\'a}bor Ivanyos, Youming Qiao, and KV~Subrahmanyam.
\newblock Non-commutative {E}dmonds' problem and matrix semi-invariants.
\newblock {\em Computational Complexity}, 26(3):717--763, 2017.
\newblock \href {http://arxiv.org/abs/1508.00690} {\path{arXiv:1508.00690}}.

\bibitem[JSV04]{jerrum2004polynomial}
Mark Jerrum, Alistair Sinclair, and Eric Vigoda.
\newblock A polynomial-time approximation algorithm for the permanent of a
  matrix with nonnegative entries.
\newblock {\em Journal of the ACM (JACM)}, 51(4):671--697, 2004.

\bibitem[KI04]{KabImp}
Valentine Kabanets and Russell Impagliazzo.
\newblock Derandomizing polynomial identity tests means proving circuit lower
  bounds.
\newblock {\em Computational Complexity}, 13:1--46, 2004.

\bibitem[KK96]{KALANTARI199687}
Bahman Kalantari and Leonid Khachiyan.
\newblock On the complexity of nonnegative-matrix scaling.
\newblock {\em Linear Algebra and its Applications}, 240:87 -- 103, 1996.
\newblock URL:
  \url{http://www.sciencedirect.com/science/article/pii/002437959400188X},
  \href {http://dx.doi.org/https://doi.org/10.1016/0024-3795(94)00188-X}
  {\path{doi:https://doi.org/10.1016/0024-3795(94)00188-X}}.

\bibitem[KLLR17]{KLLR17}
Tsz~Chiu Kwok, Lap~Chi Lau, Yin~Tat Lee, and Akshay Ramachandran.
\newblock The {Paulsen} problem, continuous operator scaling, and smoothed
  analysis.
\newblock {\em arXiv preprint arXiv:1710.02587}, 2017.

\bibitem[KN79]{KN79}
George Kempf and Linda Ness.
\newblock The length of vectors in representation spaces.
\newblock In {\em Algebraic Geometry, Lecture Notes in Math.}, pages 233--243.
  1979.

\bibitem[Kru37]{K37}
J~Kruithof.
\newblock Telefoonverkeersrekening.
\newblock {\em De Ingenieur}, 52:E15--E25, 1937.

\bibitem[Lan15]{Landsberg2015}
J.~M. Landsberg.
\newblock Geometric complexity theory: an introduction for geometers.
\newblock {\em ANNALI DELL'UNIVERSITA' DI FERRARA}, 61(1):65--117, May 2015.
\newblock URL: \url{https://doi.org/10.1007/s11565-014-0202-7}, \href
  {http://dx.doi.org/10.1007/s11565-014-0202-7}
  {\path{doi:10.1007/s11565-014-0202-7}}.

\bibitem[Lie90]{Lieb90}
Elliot Lieb.
\newblock Gaussian kernels have only {Gaussian} maximizers.
\newblock {\em Inventiones Mathematicae}, 102:179--208, 1990.

\bibitem[Lov89]{Lov1989}
Laszlo Lovasz.
\newblock Singular spaces of matrices and their application in combinatorics.
\newblock {\em Bulletin of the Brazilian Mathematical Society}, 20:87--99,
  1989.

\bibitem[LSW98]{LSW}
Nati Linial, Alex Samorodnitsky, and Avi Wigderson.
\newblock A deterministic strongly polynomial algorithm for matrix scaling and
  approximate permanents.
\newblock {\em STOC}, pages 644--652, 1998.

\bibitem[MS02]{Mulmuley_Sohoni:2002}
Ketan~D. Mulmuley and Milind Sohoni.
\newblock Geometric complexity theory i: An approach to the p vs. np and
  related problems.
\newblock {\em SIAM J. Comput.}, 31(2):496--526, February 2002.
\newblock URL: \url{https://doi.org/10.1137/S009753970038715X}, \href
  {http://dx.doi.org/10.1137/S009753970038715X}
  {\path{doi:10.1137/S009753970038715X}}.

\bibitem[Mum65]{Mum65}
David Mumford.
\newblock {\em Geometric invariant theory}.
\newblock Springer-Verlag, Berlin-New York, 1965.

\bibitem[NM84]{ness1984stratification}
Linda Ness and David Mumford.
\newblock A stratification of the null cone via the moment map.
\newblock {\em American Journal of Mathematics}, 106(6):1281--1329, 1984.
\newblock \href {http://dx.doi.org/10.2307/2374395}
  {\path{doi:10.2307/2374395}}.

\bibitem[RS89]{ROTHBLUM1989737}
Uriel~G. Rothblum and Hans Schneider.
\newblock Scalings of matrices which have prespecified row sums and column sums
  via optimization.
\newblock {\em Linear Algebra and its Applications}, 114-115:737 -- 764, 1989.
\newblock Special Issue Dedicated to Alan J. Hoffman.
\newblock URL:
  \url{http://www.sciencedirect.com/science/article/pii/0024379589904916},
  \href {http://dx.doi.org/https://doi.org/10.1016/0024-3795(89)90491-6}
  {\path{doi:https://doi.org/10.1016/0024-3795(89)90491-6}}.

\bibitem[RS05]{RazShp}
Ran Raz and Amir Shpilka.
\newblock Deterministic polynomial identity testing in non commutative models.
\newblock {\em Computational Complexity}, 14:1--19, 2005.

\bibitem[SdB01]{SVdB2001}
Aidan Schofield and Michel~Van den Bergh.
\newblock Semi-invariants of quivers for arbitrary dimension vectors.
\newblock {\em Indagationes Mathematicae}, 12(1):125--138, 2001.

\bibitem[Sin64]{Sink}
R.~Sinkhorn.
\newblock A relationship between arbitrary positive matrices and doubly
  stochastic matrices.
\newblock {\em The Annals of Mathematical Statistics}, 35:876--879, 1964.

\bibitem[Sto62]{S62}
Richard Stone.
\newblock Multiple classifications in social accounting.
\newblock {\em Bulletin de l'Institut International de Statistique},
  39(3):215--233, 1962.

\bibitem[Stu08]{sturmfels2008algorithms}
Bernd Sturmfels.
\newblock {\em Algorithms in invariant theory}.
\newblock Springer, 2008.

\bibitem[SZ90]{SZ90}
Michael~H Schneider and Stavros~A Zenios.
\newblock A comparative study of algorithms for matrix balancing.
\newblock {\em Operations research}, 38(3):439--455, 1990.

\bibitem[Tao16]{T16}
Terence Tao.
\newblock A symmetric formulation of the
  {Croot}-{Lev}-{Pach}-{Ellenberg}-{Gijswijt} capset bound.
\newblock Blog post, 2016.
\newblock
  \href{https://terrytao.wordpress.com/2016/05/18/a-symmetric-formulation-of-the-croot-lev-pach-ellenberg-gijswijt-capset-bound}{https://terrytao.wordpress.com/2016/05/18/a-symmetric-formulation-of-the-croot-lev-pach-ellenberg-gijswijt-capset-bound}.

\bibitem[Val79a]{V79comp-algebra}
Leslie~G Valiant.
\newblock Completeness classes in algebra.
\newblock In {\em Proceedings of the eleventh annual ACM symposium on Theory of
  computing}, pages 249--261. ACM, 1979.

\bibitem[Val79b]{V79perm-comp}
Leslie~G Valiant.
\newblock The complexity of computing the permanent.
\newblock {\em Theoretical computer science}, 8(2):189--201, 1979.

\bibitem[VDDM03]{verstraete2003normal}
Frank Verstraete, Jeroen Dehaene, and Bart De~Moor.
\newblock Normal forms and entanglement measures for multipartite quantum
  states.
\newblock {\em Physical Review A}, 68(1):012103, 2003.
\newblock \href {http://arxiv.org/abs/quant-ph/0105090}
  {\path{arXiv:quant-ph/0105090}}, \href
  {http://dx.doi.org/10.1103/PhysRevA.68.012103}
  {\path{doi:10.1103/PhysRevA.68.012103}}.

\bibitem[Woo11]{Woodward11}
Chris Woodward.
\newblock Moment maps and geometric invariant theory.
\newblock {\em Lecture notes,
  \href{https://arxiv.org/abs/0912.1132}{arXiv:0912.1132}}, 2011.

\end{thebibliography}

\end{document}